\documentclass[a4paper,10pt]{article}
\pdfoutput=1
\usepackage{jheppub}

\usepackage[utf8]{inputenc}
\usepackage[T1]{fontenc}
\usepackage[english]{babel}
\usepackage{fancyhdr}
\usepackage{booktabs}
\usepackage{hyperref} 
\usepackage{amssymb,amsfonts,amsmath}
\usepackage{graphics,graphicx,pstricks,color,colortbl}
\usepackage{slashed}
\usepackage{cleveref}
\crefname{figure}{Figure}{Figures}
\usepackage{subcaption}
\usepackage{multirow}
\usepackage{booktabs}
\usepackage{axodraw2}
\usepackage{braket}
\usepackage{amsmath}
\usepackage{comment}
\usepackage{graphicx}
\usepackage{tikz}
\usepackage{xcolor}
\usepackage{float}
\usepackage[T1]{fontenc}
\usepackage{placeins}
\usepackage{amsfonts}
\usepackage{mathrsfs}
\usepackage{bm}
\usepackage[normalem]{ulem}
\usepackage{orcidlink}
\usepackage{tabularx}

\newcommand{\fb}{{\rm\ fb}}
\newcommand{\GeV}{{\rm\ GeV}}
\newcommand{\TeV}{{\rm\ TeV}}

\usepackage{subcaption} 
\usepackage[english]{babel} 
\usepackage{gensymb} 
\usepackage{booktabs} 
\usepackage{nicefrac} 
\usepackage{listings} 
\usepackage{float}  
\usepackage{siunitx} 
\usepackage{comment} 
\usepackage{indentfirst} 
\usepackage[most]{tcolorbox} 

\title{Resonant Di-Higgs Searches in $b\bar b\tau^+\tau^-$ at HL-LHC: 
Supersymmetry versus Compositeness Benchmarks}

\author[a]{Stefania De Curtis\,\orcidlink{XXXX-XXXX-XXXX-XXXX},}
\author[b]{Luigi Delle Rose\,\orcidlink{XXXX-XXXX-XXXX-XXXX},}
\author[c]{Atri Dey\,\orcidlink{0000-0002-1645-7641},}
\author[c]{Carl Johan Königsson\,\orcidlink{XXXX-XXXX-XXXX-XXXX},}
\author[c, d]{Stefano Moretti\,\orcidlink{0000-0002-8601-7246},}
\author[b]{Luca Panizzi\,\orcidlink{XXXX-XXXX-XXXX-XXXX}}

\affiliation[a]{INFN, Sezione di Firenze \& Dipartimento di Fisica e Astronomia, Università di Firenze, Via G. Sansone 1, Sesto Fiorentino, 50019 Firenze, Italy}
\affiliation[b]{Dipartimento di Fisica, Universit\`a della Calabria \& 
INFN, Gruppo Collegato di Cosenza, Arcavacata di Rende, 87036, Cosenza, Italy}
\affiliation[c]{Department of Physics and Astronomy, Uppsala University,
Box 516, 751 20 Uppsala, Sweden}
\affiliation[b]{School of Physics and Astronomy, University of Southampton, Highfield, Southampton SO17 1BJ, United Kingdom}

\emailAdd{atri.dey@physics.uu.se}
\emailAdd{carljohan.konigsson.0852@student.uu.se}
\emailAdd{stefano.moretti@cern.ch}

\abstract{We explore the scope of the High-Luminosity Large Hadron Collider (HL-LHC) in testing Standard Model (SM) di-Higgs production and decay into $2b2\tau$ final states in the context of two viable theories of the Electro-Weak (EW) scale: Supersymmetry and Compositeness. Specifically, we target minimal model realisations of these two scenarios that enable the resonant process $gg\to H\to hh\to b\bar b \tau^+\tau^-$, where $h$ is the SM-like Higgs state and $H$ a heavier CP-even companion,
with the $\tau$'s decaying hadronically. This is realised within the Next-to-Minimal Supersymmetric SM  and the Composite 2-Higgs Doublet Model, respectively. Using two illustrative benchmark points, after performing a thorough detector-level Monte Carlo (MC) analysis exploiting new observables, we show that, in the former case, there is more moderate sensitivity, owing to large backgrounds, whereas, in the latter case, substantial scope for discovery exists, in a background free environment.}


\begin{document}

\maketitle

\section{Introduction}

The discovery of a spin-0 neutral CP-even (scalar) particle with a mass near 125 \GeV\ by the ATLAS and CMS collaborations at the Large Hadron Collider (LHC) constitutes one of the most important achievements in particle physics, as it confirms the mechanism of EW Symmetry Breaking (EWSB) predicted within the SM \cite{ATLAS:2012yve,CMS:2012qbp}. 
In fact, since then, increasingly precise measurements of the properties of the observed {particle} have shown remarkable agreement with {the} SM {Higgs boson} expectations~\cite{ATLAS:2022vkf, ATLAS:2022yrq, ATLAS:2015yey, CMS:2022dwd}. Nevertheless, the Higgs sector remains one of the least experimentally constrained parts of the SM, and several open questions still point towards the existence of physics Beyond the SM (BSM). Among these are, for example, the hierarchy problem emerging from interactions of the Higgs state with the top quark, which is of a theoretical nature, plus the origin of dark matter and the baryon asymmetry of the Universe, both of which are of experimental origin. Indeed, the possibility of a richer Higgs sector than the minimal realisation present in the SM, in line with the multitude of both matter states and force carriers, could offer a common handle to address all such flaws.

In this context, SM di-Higgs (or Higgs pair) production has emerged as one of the most promising probes of a wider Higgs sector. In fact, it offers direct access to the Higgs trilinear self-coupling and therefore to the structure of the Higgs potential responsible for EWSB. Within the SM, however, the di-Higgs production cross section remains very small due to the destructive interference between the triangle and box contributions in the dominant gluon-gluon Fusion (ggF) production mechanism. As a consequence, observing SM
di-Higgs production at the LHC is extremely challenging, even in the high-luminosity phase of the machine (HL-LHC) \cite{Gianotti:2002xx}. Any significant enhancement in the 
di-Higgs production rate would therefore represent a strong indication of BSM dynamics.

A generic feature of many BSM scenarios with extended Higgs sectors is the possibility of resonant di-Higgs production through the production and decay of a heavier scalar state into a pair of SM Higgs bosons\footnote{Herein, we refer to the Higgs boson discovered in 2012 as the SM Higgs boson.},
\[
gg \to H \to hh.
\]
Such resonant contributions can enhance the di-Higgs rate well beyond the SM expectation and considerably improve the discovery prospects at the HL-LHC. Furthermore, the kinematic properties of the di-Higgs system can carry important information regarding the nature of the underlying Higgs sector and the ensuing dynamics responsible for EWSB.

Among the various BSM realisations, Supersymmetry frameworks are very elegant in solving the hierarchy problem through the existence of top-quark companion states (so-called stops) while also requiring extended Higgs sectors. Among these, the Next-to-Minimal Supersymmetric SM (NMSSM) is one of the most widely studied and phenomenologically richer scenarios. Indeed, the NMSSM extends the Minimal Supersymmetric SM (MSSM) by introducing an additional gauge-singlet superfield for a purpose of alleviating the so-called $\mu$-problem while maintaining the attractive features of Supersymmetry \cite{Ellwanger:2009dp,Dedes:2000jp,Panagiotakopoulos:2000wp} (see also Ref.~\cite{Moretti:2019ulc}). Unlike the MSSM, where bounds from single-Higgs production constrain the heavy CP-even scalar to be above the TeV~\cite{ATLAS:2020zms,CMS:2022goy}, the
extended Higgs sector of the NMSSM can also allow for light CP-even scalars and lead to resonantly enhanced di-Higgs production.
Composite Higgs Model (CHM) scenarios provide another appealing framework to remedy the hierarchy problem. Here, the Higgs boson emerge as a pseudo-Nambu-Goldstone Boson (pNGB) originating from a new strongly interacting sector at a high energy scale~\cite{Kaplan:1983sm,Dugan:1984hq,Agashe:2004rs,Contino:2010rs}   
(see also Refs.~\cite{DeCurtis:2011yx,Panico:2015jxa}). Such constructions thus predict the Higgs boson 
as bound state alongside the presence of top-quark companions (so-called Heavy Top-Partners (HTPs)). Depending on the underlying symmetry, CHMs can also naturally accommodate additional heavy scalar resonances \cite{Mrazek:2011iu,Bertuzzo:2012ya}.     In particular, the Composite 2-Higgs Doublet Model (C2HDM) describes the two Higgs doublets as pNGBs \cite{DeCurtis:2018zvh}, whose properties are 
relevant for resonant di-Higgs production and decay.

Other than enabling the production of the $H$ state necessary for resonant $hh$ decays, both the NMSSM and C2HDM also provide additional contributions to the di-Higgs cross section in the form of new (bubble, triangle and box) loops of stops and HTPs, respectively, which can interfere with the $gg\to H\to hh$ amplitude (as well as those of SM origin).  As a result, the dynamics of resonant di-Higgs production and decay signatures in the NMSSM and C2HDM can differ substantially from those encountered in more conventional Higgs sector extensions, wherein the additional particles are solely Higgs states. Furthermore, the different interplay between modified Higgs trilinear self-couplings, additional heavy scalar states, and top-quark companions makes resonant di-Higgs production a particularly interesting probe of these two different scenarios of the EW scale.

Hence, 
in the spirit of Ref.~\cite{DeCurtis:2018iqd}, wherein the study of single-Higgs production and decay processes (amongst others) was invoked to distinguish between possible realisation of Supersymmetry and Compositeness, here, we perform a 
 comparative study of resonant di-Higgs production and decay in both the 
 NMSSM and C2HDM. This is phenomenologically particularly well motivated, as
 the mentioned differences originating from fundamentally distinct theoretical principles can possibly  reflect in differing experimental observables (typically, invariant and/or transverse masses) associated with resonant di-Higgs production and decay, in turn providing valuable insights into the dynamics of EWSB realised in these two frameworks. 
 
Among the various di-Higgs decay channels, the $2b2\tau$ final state constitutes one of the most promising signatures at the HL-LHC. While the $h\to b\bar b$ decay mode benefits from the largest Branching Ratio (BR) of the SM Higgs boson, fully hadronic final states such as $4b$ suffer from overwhelming QCD backgrounds. The inclusion of a $\tau^+\tau^-$ pairs in the $hh$ decay chain considerably improves the signal-to-background discrimination while still
maintaining a sizeable signal rate. Consequently, the $2b2\tau$ topologies offers an attractive balance between large BRs and experimental cleanliness. In addition, ongoing developments in boosted object reconstruction and $\tau$-tagging techniques are expected to further improve the sensitivity of this channel during the HL-LHC era.

Motivated by these considerations, we perform a detailed collider study of resonant di-Higgs production and decay in the $2b2\tau$ final state at the HL-LHC within the frameworks of the NMSSM and C2HDM. We consider a representative Benchmark Point (BP)  in both cases consistent with current theoretical and experimental constraints and producing {approximately} the same cross section within theoretical uncertainties (due to Parton Distribution Functions (PDFs), renormalisation/factorisation scale dependence, etc.), albeit with different mass spectra as typical in each scenario, allowing us to investigate the discovery prospects for heavy scalar resonances decaying into SM di-Higgs intermediate states. The selected BPs incorporate the key features of the two scenarios and the performed analysis is specifically tailored to highlight the peculiarities arising from the distinct particles involved (apart from $H$ state), as well as from the different mass spectrum properties, which in turn lead to distinctive contributions to the interferences.

The remainder of this paper is organised as follows. In Sec.~\ref{sec:Theory}, we briefly review the processes generating di-Higgs production in our BSM frameworks (the NMSSM in 
Subsec.~\ref{sec:NMSSM} and the C2HDM in \ref{sec:C2HDM}), including defining our Benchmark Points (BPs). In Sec.~\ref{sec:collider}, we describe the collider setup, relevant signal and background processes, event generation framework, and analysis strategy, as well as present our numerical results and projected sensitivities at the
HL-LHC. Finally, we summarise and conclude in Sec.~\ref{sec:summa}.

\section{Theoretical Frameworks}
\label{sec:Theory}
\subsection{Next-to-Minimal Supersymmetric SM (NMSSM)}
\label{sec:NMSSM}
\begin{figure}[!t]
    \centering
    \includegraphics[width=0.9\linewidth]{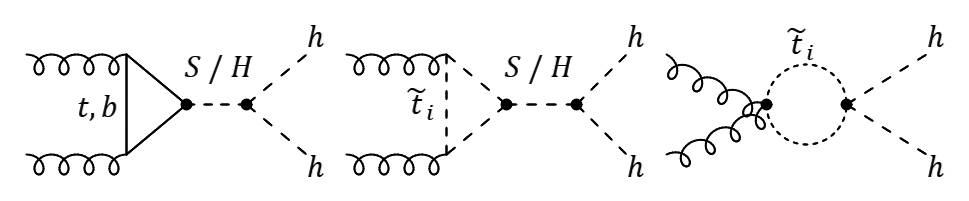}
    \caption{Representative partonic topologies contributing to the di-Higgs production in the NMSSM. The loop fermions $t, b$ are the SM bottom and top quarks while the loop scalars ${\tilde{t}}_i$ include two additional stops. Here, $H$ represents the resonant heavy scalar Higgs while ${S}'$ refers to both the SM Higgs $h$ and additional Higgs singlet $S$ of the NMSSM.}
    \hskip-10.5cm\vskip-4.85cm{~~~~\,\,'~~~~~~~~~~~~~~~~~~~~~~~~~~~~~~~~~~~~~\,'~~~~~~~~~~~~~~~~~~~~~~~~~~~~~~~~~}
    \vskip+4.25cm 
    \label{fig:nmssm_feynman}
\end{figure}
The NMSSM implementation adopted here is described in detail in Ref.~\cite{Moretti:2025dfz}.
Some of the contributing channels to di-Higgs production in the NMSSM framework can be seen in Fig. \ref{fig:nmssm_feynman}. In addition to the SM channels, two resonant channels open up through the addition of the new singlet scalar $S$ and the extra heavy Higgs boson $H$. These resonances can be reached through SM quark interactions creating $S$ and $H$, the addition of coloured squarks from the NMMSM also opens up additional channels enabling the $S$ and $H$ resonances as well as direct di-Higgs production through quartic vertices (there are also box diagrams contributing to the latter, see \cite{Moretti:2025dfz}).
\begin{table}[t]
\centering
\caption{NMSSM BP used for the resonant di-Higgs
analysis. Here, in particular, $m_H$ denotes the mass of the doublet heavy Higgs, $m_S$ denotes the mass of the singlet heavy Higgs whereas $m_{{\tilde t}_i}$ denotes the mass of the two stops.}
\label{tab:NMSSM_benchmark}
\begin{tabular}{lc}
\toprule
Parameter & Value \\
\midrule
$m_H$ [GeV] & {809} \\
$m_S$ [GeV] & {1193} \\
$\tan\beta$ & {2.32} \\
$\lambda$ & {0.73} \\
$\kappa$ & {0.73} \\
$A_\lambda$ [GeV] & {$-308.6$} \\
\midrule
$m_{\tilde t_1}$ [GeV] & {619} \\
$m_{\tilde t_2}$ [GeV] & {1391} \\
\midrule
{{$\Gamma_H/M_H$}} & {{0.73\%}} \\
{{$\sigma_{\rm NMSSM}/\sigma_{\rm SM}$}} & {{1.5}} \\
\bottomrule
\end{tabular}
\end{table}
The BP used in this paper is listed in Tab.~\ref{tab:NMSSM_benchmark}. It is a slight modification of BP4 from \cite{Moretti:2025dfz} (see Fig. 5 therein) and features an enhanced value of $\lambda_{hhh}$ and a squark contribution which is rather small, leading to negligible threshold effects at $2m_{{\tilde t}_1}$. Furthermore, both heavy scalars (the doublet $H$ and the singlet $S$) produce a resonant peak, and there is
constructive interference in the interval $m_H < m_{hh} < m_S$ together with a destructive interference over the range $m_{hh} < m_H$ and
$m_{hh} > m_S$. The constructive interference is the largest one, and it is responsible for the total cross section of this BP being larger than the SM one, which is the irreducible background. While with these parameters the di-Higgs  cross section at Leading Order (LO)  is typically around 30\% larger than the SM one, we assume that combining uncertainties from PDFs and higher-order effects would enable it to reach the 50\% level, without significantly modifying the shapes of differential observables.

\subsection{Composite 2-Higgs Doublet Model (C2HDM)} 
\label{sec:C2HDM}

\begin{figure}[!b]
\centering
\includegraphics[width=0.95\textwidth]{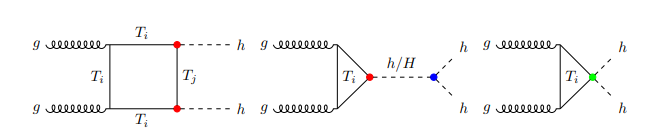}
\caption{Representative partonic topologies contributing to di-Higgs production in the C2HDM. The fermions $T_i$, $i=1,..,9$ include the SM top quark together with the eight
additional HTPs. 
}
\label{fig:C2HDM_dihiggs_diagrams}
\end{figure}

The C2HDM adopted here is described in of Ref.~\cite{DeCurtis:2023pus}, where we refer the reader for all details.
 The dominant partonic contributions to
di-Higgs production are seen in Fig.~\ref{fig:C2HDM_dihiggs_diagrams}: here, 
the presence of the effective $T_iT_ihh$ vertex (where $T_i$, $i=1,..,9$ includes the SM top quark together with the eight
additional HTPs)  is especially interesting since it generates loop structures absent in ordinary renormalisable elementary 2HDMs. 
%
The interplay among the three topologies in Fig.~\ref{fig:C2HDM_dihiggs_diagrams} can significantly modify the invariant mass spectrum of the Higgs pair system.  Furthermore,
threshold effects near
$
m_{hh}\sim2m_{T_i}
$
can appear.
%
%
%
%
For the collider analysis we select one representative resonant BP of the C2HDM, which numerical input parameters are summarised in Tab.~\ref{tab:C2HDM_benchmark}. This is BP3 of 
\cite{DeCurtis:2023pus} (see Fig.~11 therein), which 
features a resonance for the heavy Higgs mass $m_H \approx 1.2$ \TeV\ and also a threshold effect due to the HTPs setting in at $\sqrt{s}\approx  2.7$ \TeV. Due
to the relatively large Higgs mass, though, the effect of the resonance contribution on the total cross
section is rather small, so that we obtain a value for it that is again some 50\% higher than the irreducible background given by  SM di-Higgs production.
The shape of the differential cross section, however, is rather interesting since there exist several 
contributions competing against each other, stemming from the heavy resonance, the heavy top-partners and the quartic
scalar-scalar-fermion-fermion diagram, which can all interfere with each other. 
\begin{table}[!t]
\centering
\caption{C2HDM BP point used for the resonant di-Higgs
analysis. Here, 
$m_H$ denotes the mass of the heavy Higgs whereas $m_{T_i}$ denote the mass of the 
eight HTPs.
}
\label{tab:C2HDM_benchmark}
\begin{tabular}{lc}
\toprule
Parameter & Value \\
\midrule
$f$ [GeV] & {796} \\
$m_H$ [GeV] & {1182} \\
$\sin \theta$ & {0.024} \\
\midrule
$m_{T_1}$ [GeV] & {1359} \\
$m_{T_2}$ [GeV] & {1584} \\
$m_{T_3}$ [GeV] & {1615} \\
$m_{T_4}$ [GeV] & {1629} \\
$m_{T_5}$ [GeV] & {1973} \\
$m_{T_6}$ [GeV] & {3589} \\
$m_{T_7}$ [GeV] & {7388} \\
$m_{T_8}$ [GeV] & {7456} \\
\midrule
$\Gamma_H/M_H$ & {5.4\%} \\
$\sigma_{\rm C2HDM}/\sigma_{\rm SM}$ & {1.5} \\
\bottomrule
\end{tabular}
\end{table}

\section{Collider Analysis}
\label{sec:collider}
Having established the theoretical frameworks and associated BPs
for both the NMSSM and C2HDM, we now turn to the collider
analysis of resonant Higgs pair production and decay in the
\begin{equation}
  hh\to b\bar b\tau^+\tau^-  
\end{equation}
final state at the HL-LHC. The purpose of this study is not only to
investigate the observability of such  di-Higgs process  in
these BSM scenarios, but also to examine whether characteristic
differences between the two models can survive a realistic detector
environment after event reconstruction and selection.

The analysis is performed assuming proton-proton collisions at a
centre-of-mass energy
$
\sqrt{s}=14~{\rm TeV},
$
with an integrated luminosity of
$
\mathcal{L}=4000~{\rm fb}^{-1},
$
corresponding to the expected operating conditions of the
HL-LHC. Since the final state under consideration
contains both genuine QCD activity and hadronic $\tau$'s,
a realistic treatment of detector effects and tagging efficiencies
is essential in order to obtain meaningful collider sensitivities.

The hard-scattering signal and background events are generated using
{\tt MadGraph5\_aMC@NLO} \cite{Alwall:2014hca}. For both the NMSSM and C2HDM
analysis, the model implementation includes the complete loop-induced
structure relevant for resonant Higgs pair production, including the
effects of the stops and the HTPs, respectively.

Parton showering, hadronisation and unstable particle decays are
simulated using {\tt Pythia8} \cite{Pythia8, Sjostrand:2014zea} while detector
response effects are incorporated through {\tt Delphes3} \cite{Delphes2, deFavereau:2013fsa}
using an HL-LHC detector configuration. Jets are reconstructed using
the anti-$k_T$ clustering algorithm \cite{Cacciari:2008gp}, as
implemented in {\tt FastJet} \cite{Cacciari:2011ma}, with jet radius
parameter
$
R=0.4.
$

Particular emphasis is placed on the reconstruction of hadronically
decaying $\tau$ leptons. In addition to standard kinematic variables,
we also employ observables sensitive to the polarisation structure
of the $\tau$ decays. Since the visible decay products of polarised
$\tau$ leptons retain information about the helicity structure of
the parent particle, such observables can provide additional
discrimination power between signal and background processes. In the
present analysis, this information is incorporated through the
leading pion momentum fraction inside the reconstructed $\tau$-jet,
which proves especially useful in suppressing backgrounds while
retaining a significant fraction of the signal events.

All event yields quoted throughout this work are obtained after full
detector simulation and normalised to the aforementioned HL-LHC integrated
luminosity. In the following subsections, we discuss the signal and
background processes considered in the analysis, the kinematic
observables employed for event selection and the resulting
significance estimates for the NMSSM and C2HDM BPs.

\subsection{Signal and Background Processes}

As mentioned in the Introduction, the $b\bar b\tau^+\tau^-$ final
state provides an attractive compromise between signal rate and
background rejection. While the decay
$
h\to b\bar b
$
benefits from the largest BR of the SM Higgs boson,
fully hadronic channels such as $4b$ suffer from overwhelming QCD
backgrounds. The inclusion of a $\tau^+\tau^-$ pair considerably
improves the signal-to-background discrimination while still
retaining a sizeable signal yield. Furthermore, as mentioned, the hadronic
$\tau$ decays allow the use of polarisation sensitive observables,
which can provide additional discrimination between signal and
background processes.

The dominant signal topology considered throughout this analysis is
the resonant production process
$
gg\to H\to hh\to b\bar b\tau^+\tau^-,
$
where $H$ denotes the heavy scalar resonance predicted by the
corresponding BSM framework. In the NMSSM(C2HDM), the ggF
production mechanism receives additional contributions from the
stops(HTPs) discussed in previous sections, leading
to potentially sizeable modifications of both the production rate
and  kinematic structure of the Higgs pair system. SM 
di-Higgs production (an irreducible background) is also included for comparison in order to
evaluate the enhancement induced by the NMSSM and C2HDM scenarios, which then takes the form of
a resonant BSM contribution plus the interferences with the SM.

At detector level, events are required to contain at least two
identified $b$-tagged jets together with two reconstructed
$\tau$-tagged jets (which will be referred to in Secs. \ref{cut_NMSSM} and \ref{cut_C2HDM} as Pre-Cut). Since the analysis is performed after full
parton showering and detector simulation, reducible backgrounds
arising from jet mistagging become particularly important. The
following SM processes are therefore found to constitute the
dominant backgrounds to the signal.

\begin{itemize}

\item \textit{Top-antitop production:}
$
pp\to t\bar t.
$
This represents the dominant background throughout the analysis due
to its extremely large production cross section. The fully
leptonic decay chain
$
t\bar t
\to
W^+W^-b\bar b
\to
\tau^+\tau^-b\bar b
+\nu_\tau\bar\nu_\tau
$
directly reproduces the signal topology and therefore constitutes an
important irreducible background. In addition, semileptonic
$t\bar t$ events can also contribute significantly when light jets
originating from hadronic $W^\pm$ decays are misidentified as hadronic
$\tau$-jets. Since the final state contains genuine $b$-jets together
with substantial $\not\hspace*{-0.085cm} E_T$ from neutrinos, the
$t\bar t$ background remains particularly difficult to suppress.

\item \textit{$Z$~+~di-jet production:}
$
pp\to Zjj
$
(where $j$ represents a jet).
Needless to say, 
$Zjj$ production constitutes another major relevant background source,
especially through the decay chain
$
Zjj \to\tau^+\tau^- + jj.
$
In this case, the hadronic jets may either correspond to genuine
$b$-jets or arise from light-flavour jets that are subsequently
misidentified as $b$-jets after detector simulation.
Although the inclusive $Zjj$ production cross section is smaller than
that of $t\bar t$, the relatively clean leptonic structure of the
process allows a fraction of events to survive the event selection
criteria.

\item \textit{QCD multi-jet production:}
$
pp\to {\rm multi-jets}.
$
The QCD multi-jet background does not contain genuine 
$\tau$'s at parton level. Nevertheless, owing to the enormous
QCD production cross section at the LHC, even highly suppressed
mistagging probabilities can generate a substantial effective background.
In particular, events containing multiple energetic jets can enter
the signal region when two light jets are identified as $b$-jets and two
additional jets are mistagged as hadronic $\tau$ candidates.
Consequently, the QCD background provides an important test of the
robustness of the event selection strategy and the $\tau$-tagging
performance.

\end{itemize}

We reckon that the treatment of the QCD multi-jet background in an
experimental analysis is inevitably more involved than a 
MC estimate. In particular, in the ATLAS search for 
$hh\to b\bar b\tau^+\tau^-$, the multi-jet contribution in
the $\tau_{\rm had}\tau_{\rm had}$ channel is estimated using a
data-driven fake-factor method based on anti-identified
$\tau_{\rm had}$ control regions while fake-$\tau_{\rm had}$
contributions from $t\bar t$ are corrected using data-derived scale
factors~\cite{ATLAS:2022xzm}. Such a procedure is designed to
capture the $j$-to-$\tau$ misidentification rate in data and cannot
be fully reproduced in a purely MC phenomenological study.

{In the present work, QCD multi-jet samples are generated and passed
through the same detector-level selection as the signal and other
backgrounds in order to assess their possible impact. After requiring
two $b$-tagged jets and two reconstructed $\tau$-tagged jets, a
non-negligible QCD contribution remains, as expected from the large
inclusive multi-jet cross section. However, once the high-mass
resonance requirement $M(bb\tau\tau)>500~{\rm GeV}$ or the requirement for total transverse activity, $H_T > 400~{\rm GeV}$, are imposed to eliminate major SM backgrounds (see Sec.~\ref{cut_NMSSM} and Sec.~\ref{cut_C2HDM}), the
QCD contribution becomes negligible within the available simulated
statistics. Since the subsequent optimised cuts further suppress
fake-object topologies, while the dominant surviving backgrounds are
found to be $t\bar t$ and $Zjj$, we focus the final detector-level
distributions and cut-flow discussion on these backgrounds only. A fully
data-driven determination of the residual QCD fake background is
beyond the scope of the present phenomenological analysis.}

At parton level and even after parton showering, the resonant heavy
scalar contribution can often be reconstructed relatively clearly   in
the invariant mass spectrum of the Higgs pair system. However, after
including detector effects, jet reconstruction uncertainties,
neutrinos from hadronic $\tau$ decays and detector smearing, the
resonance structure becomes substantially broadened. As a result,
the reconstruction of the heavy scalar resonance becomes considerably
more challenging at detector level.

In order to recover as much resonance information as possible, we
therefore study several kinematic observables sensitive to the
underlying heavy scalar structure. In particular, the invariant mass
variable of the full final state
plays a central role in the analysis since it provides the closest
detector-level approximation to the invariant mass of the original
Higgs pair system. Additional observables involving transverse
momentum distributions, angular separations and $\tau$-polarisation
sensitive variables are also employed in order to improve the
signal-to-background discrimination.
The corresponding kinematic distributions and the resulting event
selection strategy are discussed in the next subsection.

\subsection{Analysis Strategy and Kinematic Observables}

In order to establish the resonant di-Higgs signal against the large
SM backgrounds, several detector-level kinematic
observables are investigated throughout this analysis. Since the
presence of neutrinos from hadronic $\tau$ decays together with
detector smearing effects significantly deteriorates the direct
reconstruction of the heavy scalar resonance, the analysis strategy
relies on a combination of invariant mass observables, transverse energy/momentum variables, angular separation variables and
$\tau$-polarisation-sensitive observables.


The most important observable in the present analysis is therefore
the reconstructed invariant mass variable
$
M(bb\tau\tau),
$
which provides the closest detector-level approximation to the
invariant mass of the original Higgs pair system. Specifically, since resonant
di-Higgs production is expected to populate the high-mass region,
this variable provides significant discrimination between signal and
background processes.

In addition, several
other observables are employed in the analysis. The invariant masses
$
M(bb)~{\rm and}~M(\tau\tau)
$
are used to reconstruct the two SM Higgs candidates individually. Signal
events are expected to cluster around the Higgs mass peak around 125 GeV while
background processes typically exhibit broader distributions.
Transverse momentum observables such as
$
p_T(b_1)
~{\rm and}~
p_T(\tau_1)
$
and the total hadronic activity
$
H_T
$
also provide important discrimination power. Since the heavy scalar
resonance typically produces boosted Higgs bosons, the corresponding
decay products are expected to possess harder transverse momentum
spectra compared to the dominant backgrounds.

Angular separation variables, like
$
\Delta R(bb)
~{\rm and}~
\Delta R(\tau\tau),
$
are particularly useful for identifying boosted Higgs decay
topologies. In fact, in resonant di-Higgs production, the decay products of
the Higgs bosons tend to become more collimated, leading to smaller
angular separations between the reconstructed objects.

Special emphasis is placed on the variable
$
Z_g={p_T(\pi_1)}/{p_T(\tau_1)},
$
which represents the leading pion momentum fraction inside the
hadronically decaying $\tau$-jet. This observable is sensitive to
the polarisation structure of the parent $\tau$ particle since the
visible momentum distribution of the decay products retains
information about the helicity configuration of the original decay \cite{Dey:2021sug, Dey:2021alu, De:2024puh}.
Consequently, as intimated, $Z_g$ provides additional discrimination power between
the resonant di-Higgs signal and the dominant SM backgrounds.

The kinematic observables employed throughout the analysis are
summarised in Tab.~\ref{tab:KinematicObservablesTable}.

\begin{table}[t]
\centering
\small
\renewcommand{\arraystretch}{1.25}
\begin{tabularx}{\linewidth}{
    >{\raggedright\arraybackslash}p{0.25\linewidth}
    >{\raggedright\arraybackslash}p{0.25\linewidth}
    >{\raggedright\arraybackslash}X}
\toprule
Observable & Symbol & Definition / Physical role \\
\midrule

Invariant masses
& $M(bb)$, $M(\tau\tau)$, $M(bb\tau\tau)$
& Reconstruction of the two Higgs candidates and the heavy scalar resonance \\

Transverse momentum
& $p_T(b_1)$, $p_T(\tau_1)$
& Identification of boosted Higgs decay products \\

Scalar $p_T$ sum
& $H_T$
& Total transverse activity of the event,
$H_T=\sum_i p_T^i$ \\

Missing transverse energy
& $\not\hspace*{-0.085cm} E_T$
& Sensitivity to neutrinos from hadronic $\tau$ decays \\

Angular separation
& $\Delta R(bb)$, $\Delta R(\tau\tau)$
& Probe of boosted Higgs decay topologies \\

Leading pion momentum fraction
& $Z_g=\dfrac{p_T(\pi_1)}{p_T(\tau_1)}$
& Sensitive to the polarisation of the reconstructed hadronic
$\tau$-jet \\

Invariant mass ratios
& $\dfrac{M(bb)}{M(bb\tau\tau)}$,
  $\dfrac{M(\tau\tau)}{M(bb\tau\tau)}$
& Correlation between reconstructed Higgs candidates and the heavy resonance \\

\bottomrule
\end{tabularx}

\caption{Detector-level kinematic observables employed in the
analysis. These variables are used to reconstruct the SM Higgs
candidates, probe the heavy scalar resonance structure and optimise
the signal-to-background discrimination.}
\label{tab:KinematicObservablesTable}
\end{table}

In the following subsections, normalised detector-level distributions
of these observables are presented for both the NMSSM and C2HDM
BPs together with the corresponding background
processes. Based on the behaviour of these distributions, optimised
selection criteria are constructed in order to maximise the
sensitivity to our resonant di-Higgs signal.

For a process with production cross section $\sigma$ and integrated
luminosity $\mathcal{L}$, the expected number of detector-level
events after selection cuts is computed as
\begin{equation}
N_L
=
\sigma\times\mathcal{L}
\times
\frac{N_{\rm cut}}{N_{\rm gen}},
\label{eq:NL}
\end{equation}
where $N_{\rm gen}$ and $N_{\rm cut}$ denote the number of generated
and surviving events, respectively.
The statistical significance is estimated using the Asimov
approximation~\cite{Cowan:2010js}
\begin{equation}
\mathcal{S}
=
\sqrt{
2\left[
(S+B)\ln\left(1+\frac{S}{B}\right)-S
\right]
},
\label{eq:Significance}
\end{equation}
where $S$ and $B$ denote the expected number of signal and background
events after event selection. In the limit
$
B\gg S,
$
the significance reduces to the following familiar Gaussian approximation (not used here):
\begin{equation}
\mathcal{S}
\simeq
\frac{S}{\sqrt{B}}.
\label{eq:S_to_B_ratio}
\end{equation}
{(Note that the significance formulae in the last two equations do not take into account the uncertainties on the background, which will be discussed at the end of the two sections dedicated to the BP analyses.)}

In the following subsections, the distributions of the relevant
kinematic observables are analysed and optimised selection cuts are
constructed separately for the NMSSM and C2HDM BPs.


\subsection{Detector-level Analysis of the NMSSM BP}
\label{cut_NMSSM}

We now turn to the detector-level analysis of the NMSSM BP at
the HL-LHC. At this stage, no additional kinematic selections are imposed, such
that the characteristic detector-level features of both the signal
and background processes can be examined before optimising the
event selection.

Unlike many conventional resonance searches, the challenge of the
present analysis is twofold. Besides suppressing the dominant
SM backgrounds arising from $t\bar t$ and $Zjj$
production and decay, it is equally important to distinguish the resonant
NMSSM contribution from the irreducible SM di-Higgs
production. Since both the NMSSM and the SM processes
lead to the identical final state, 
their detector signatures share many common features. Consequently,
the discrimination between the two relies primarily on the modified
production mechanism of the heavy scalar resonance, which gives rise
to more energetic Higgs bosons and a different overall event
topology than the non-resonant SM process.

\begin{figure}[htpb!]
    \centering
    \begin{subfigure}[b]{0.48\linewidth}
        \centering
        \includegraphics[width=\linewidth, height=6cm]{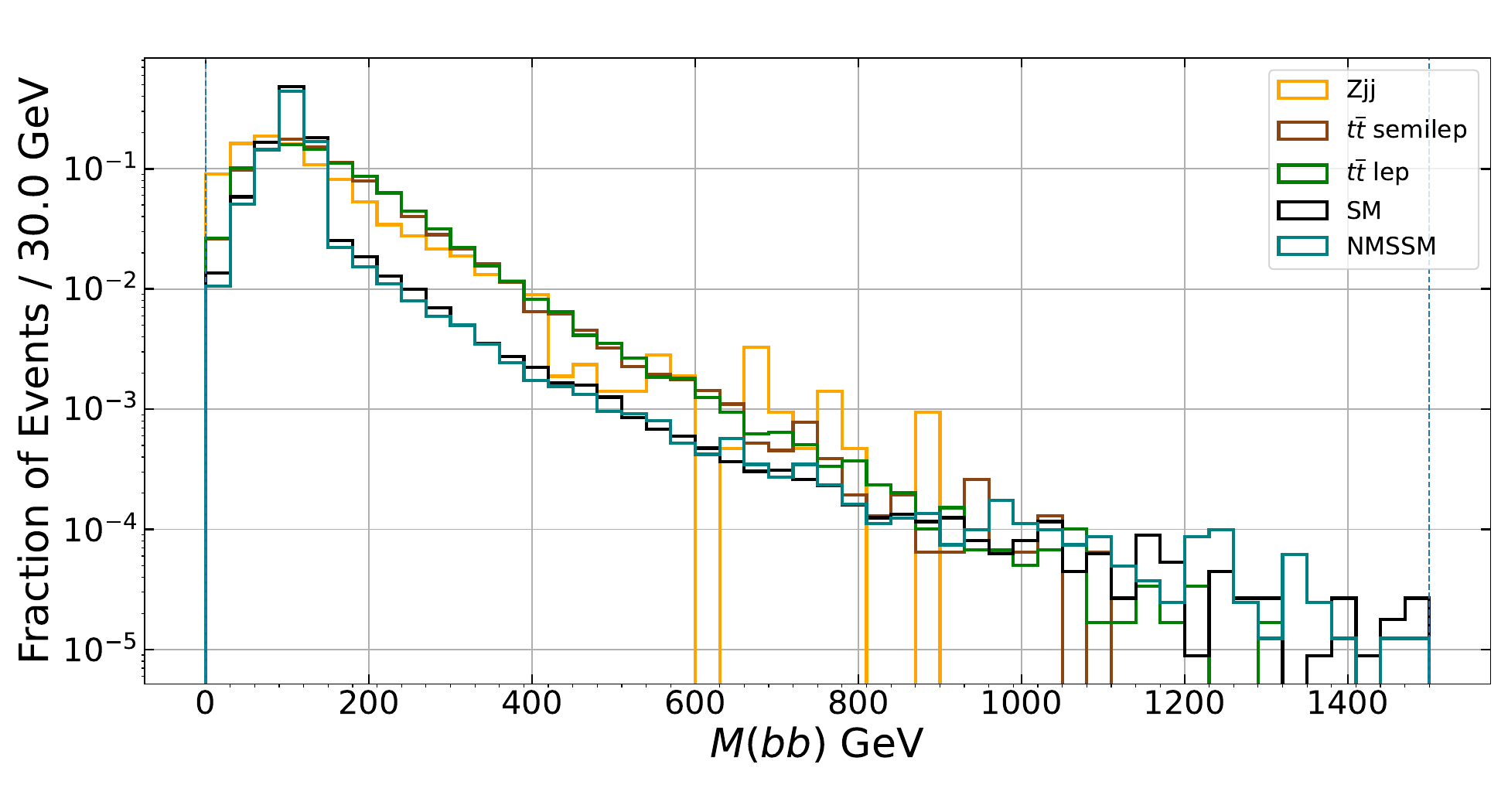}
        \caption{Invariant mass $M(bb)$.}
    \end{subfigure}
    \hfill
    \begin{subfigure}[b]{0.48\linewidth}
        \centering
        \includegraphics[width=\linewidth, height=6cm]{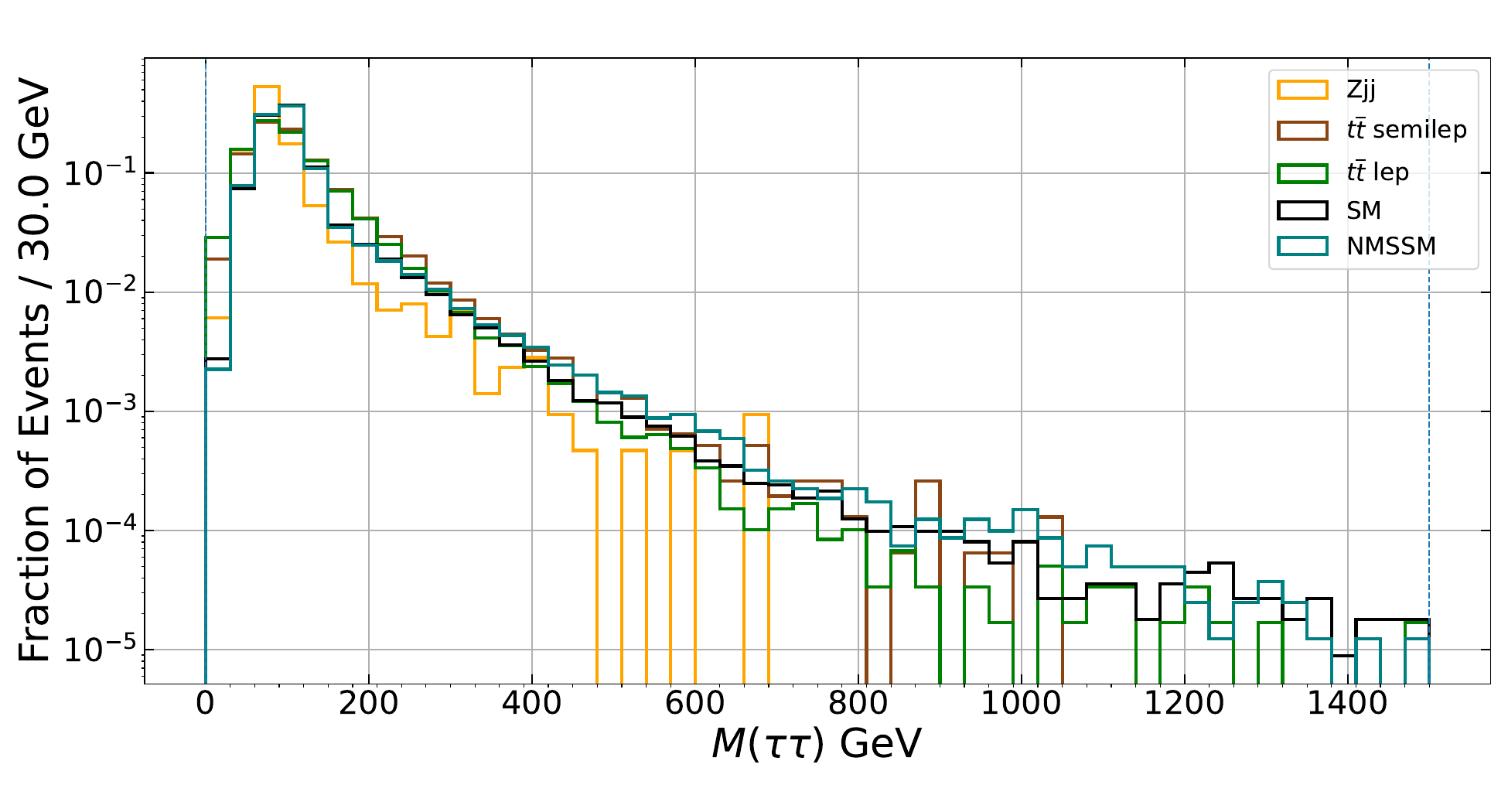}
        \caption{Invariant mass $M(\tau\tau)$.}
    \end{subfigure}
    \vspace{0.3cm}
    \begin{subfigure}[b]{0.48\linewidth}
        \centering
        \includegraphics[width=\linewidth, height=6cm]{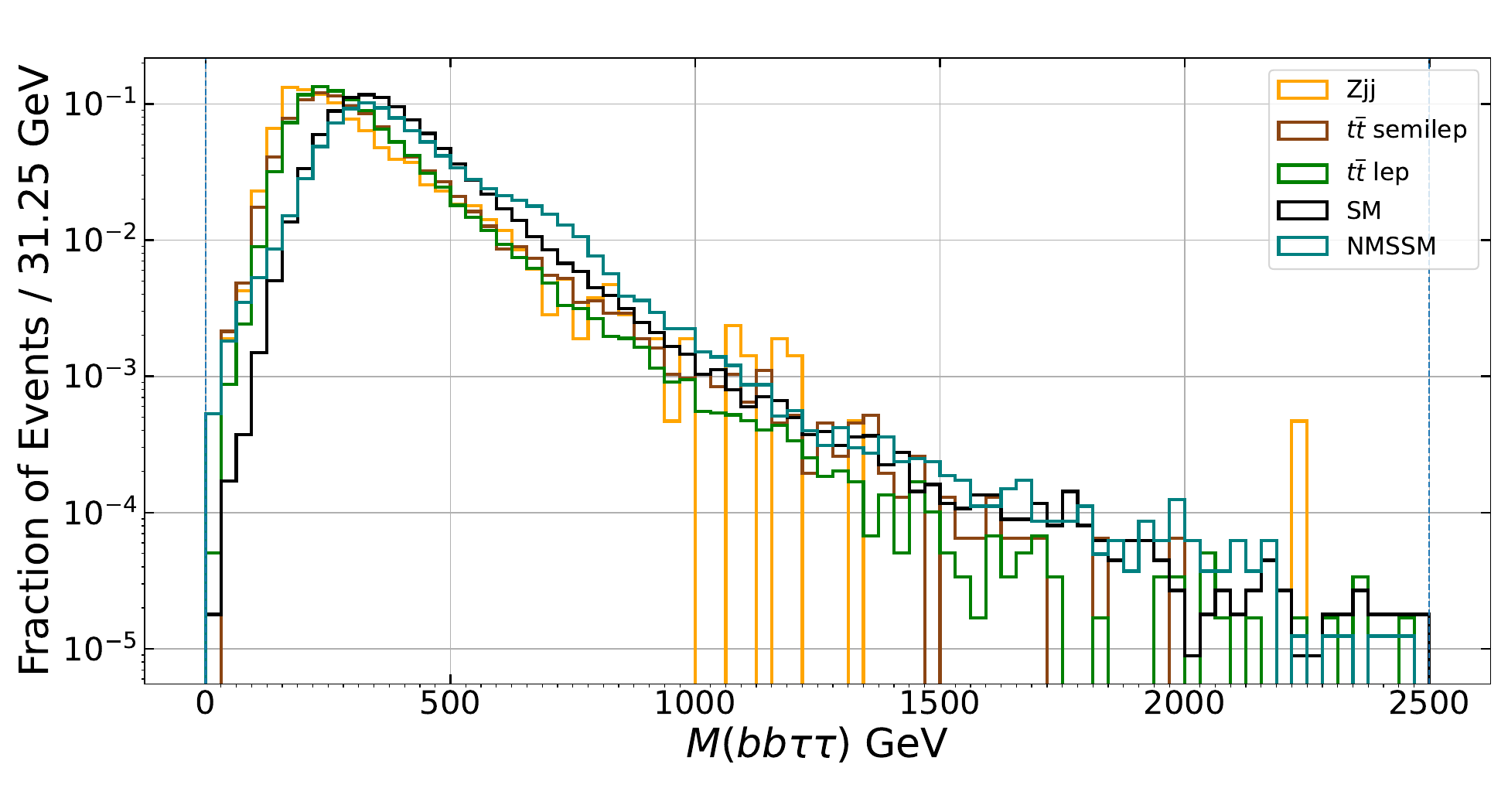}
        \caption{Invariant mass $M(bb\tau\tau)$.}
    \end{subfigure}
    \hfill
    \begin{subfigure}[b]{0.48\linewidth}
        \centering
        \includegraphics[width=\linewidth, height=6cm]{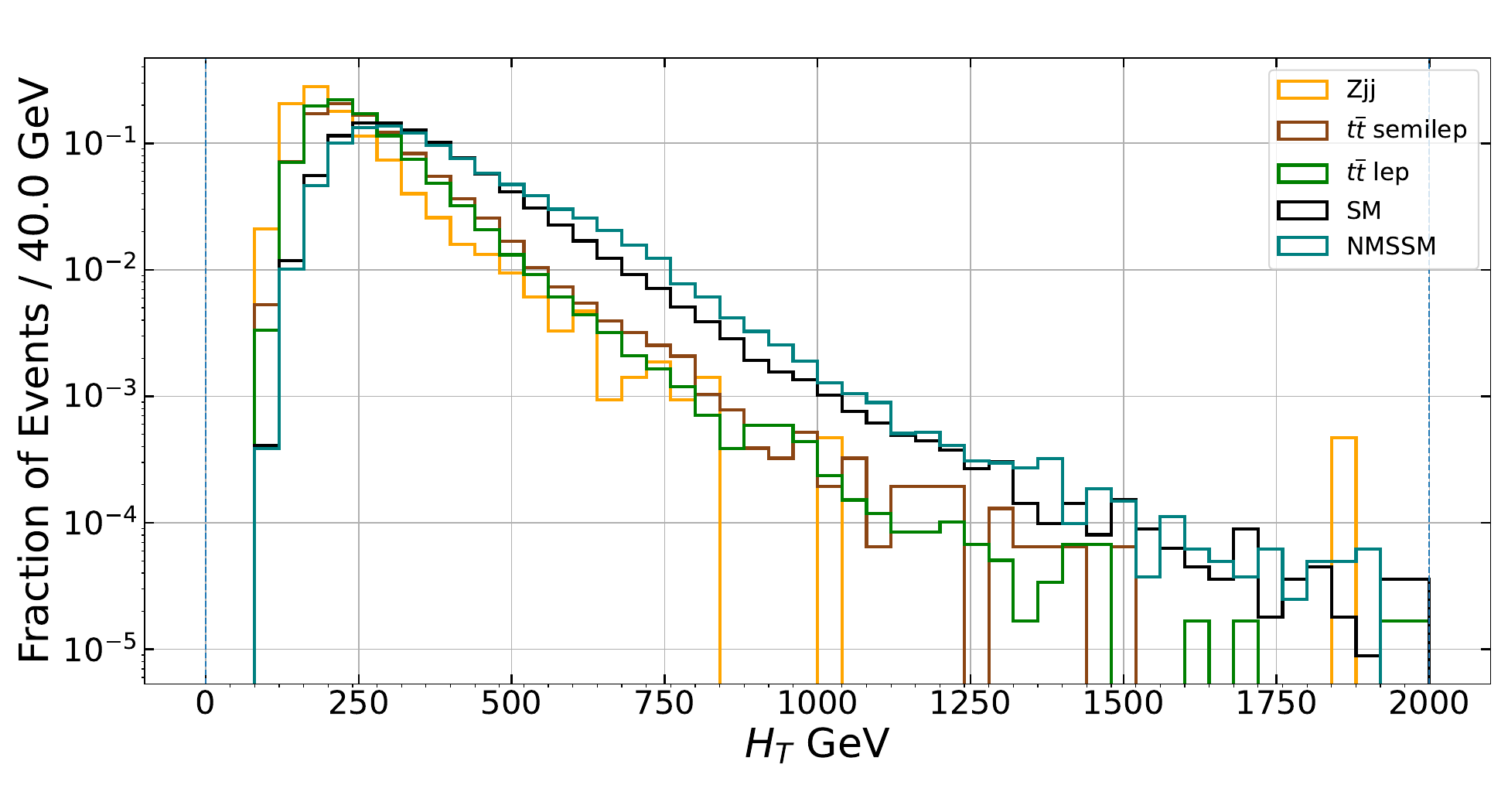}
        \caption{Transverse hadronic activity $H_T$.}
    \end{subfigure}
    \vspace{0.3cm}
    \begin{subfigure}[b]{0.48\linewidth}
        \centering
        \includegraphics[width=\linewidth, height=6cm]{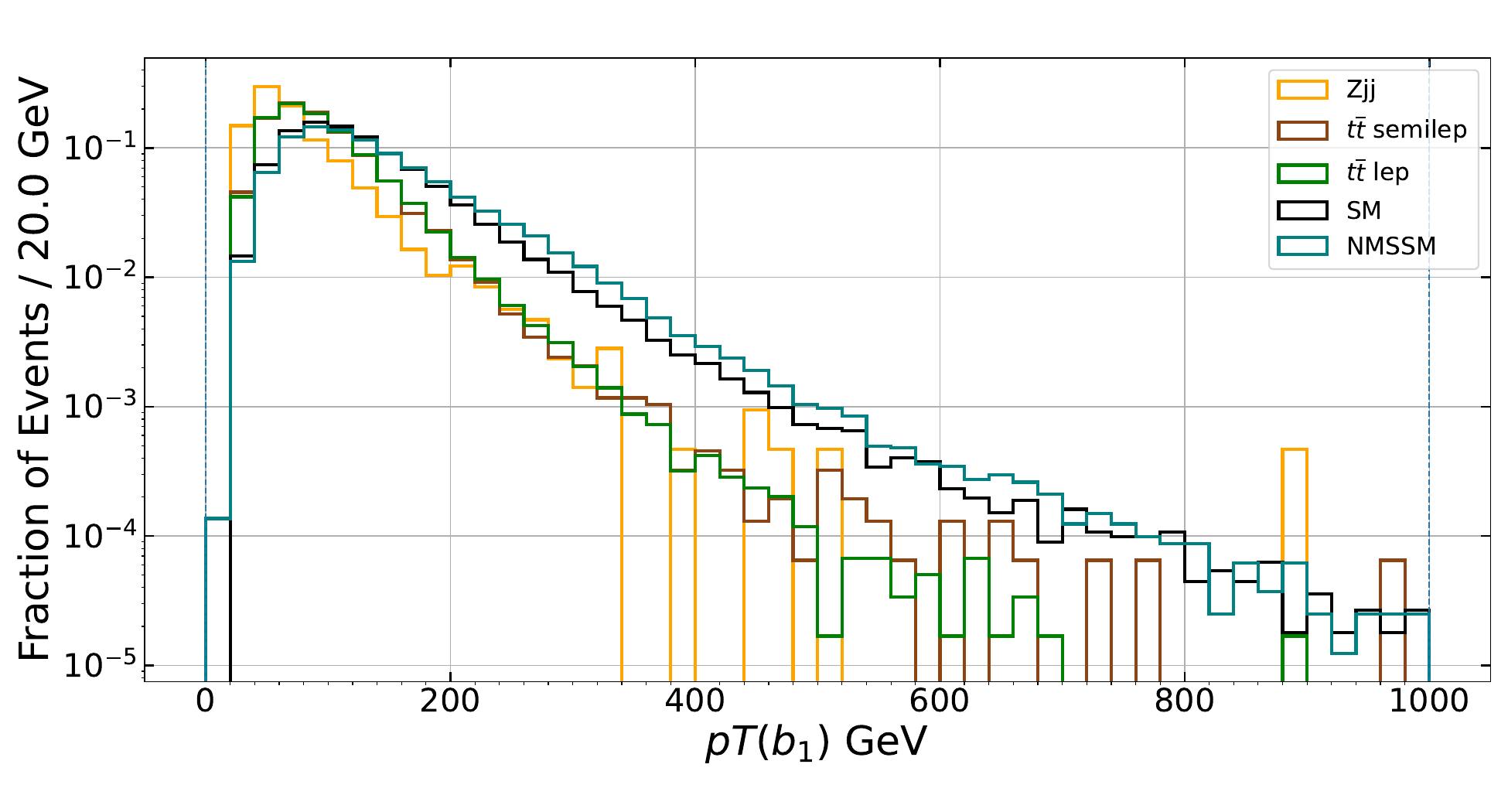}
        \caption{Transverse momentum $p_T(b_1)$.}
    \end{subfigure}
    \hfill
    \begin{subfigure}[b]{0.48\linewidth}
        \centering
        \includegraphics[width=\linewidth, height=6cm]{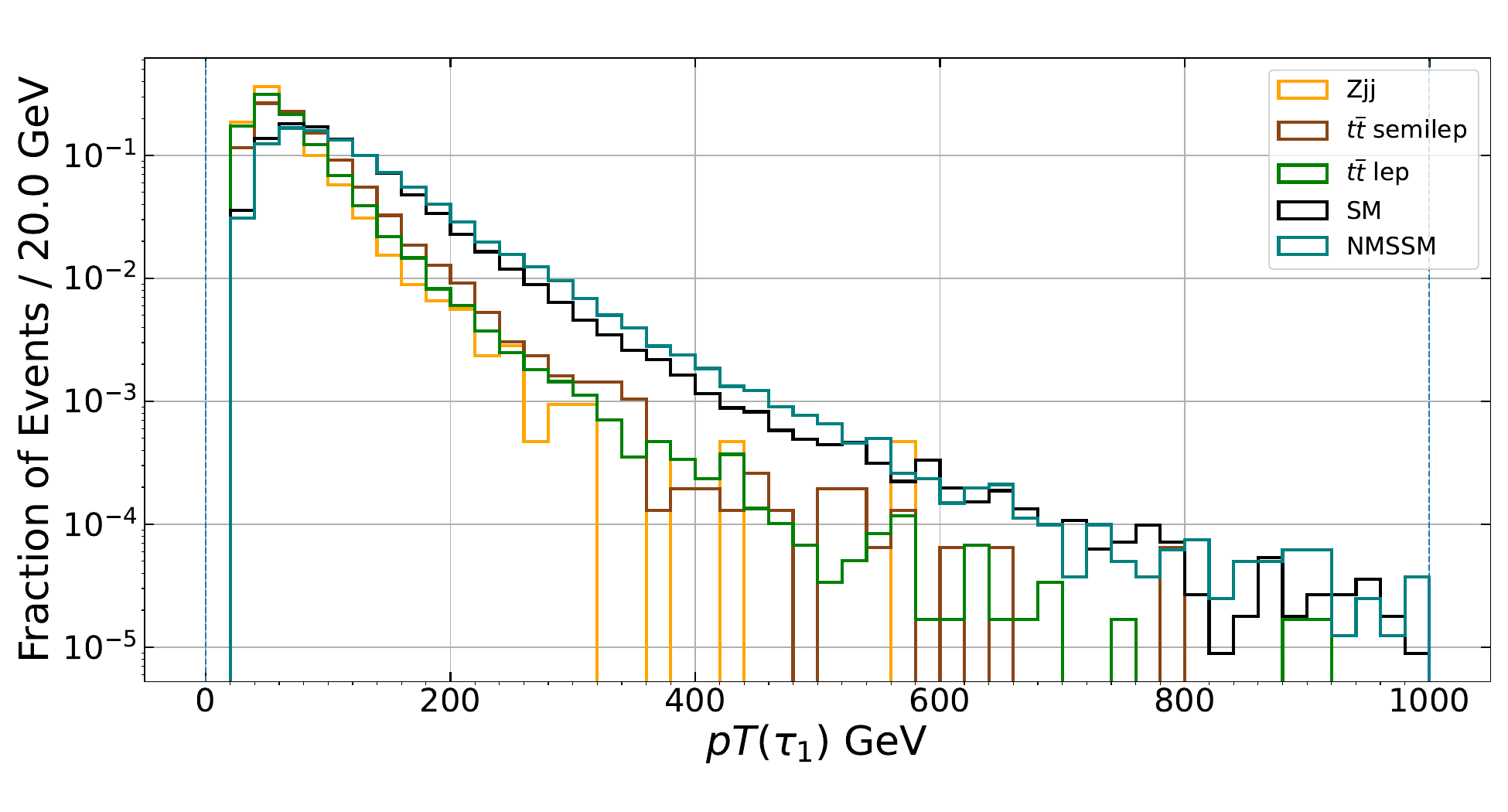}
        \caption{Transverse momentum $p_T(\tau_1)$.}
    \end{subfigure}
    \caption{Normalised distribution of reconstructed final state observables for both signals and major SM backgrounds after Pre-Cut at HL-LHC with $\sqrt{s}$ = 14 TeV for NMSSM.}
    \label{fig:NMSSM_plots1}
\end{figure}
\begin{figure}[htpb!]
    \centering

    \begin{subfigure}[b]{0.48\linewidth}
        \centering
        \includegraphics[width=\linewidth, height=6cm]{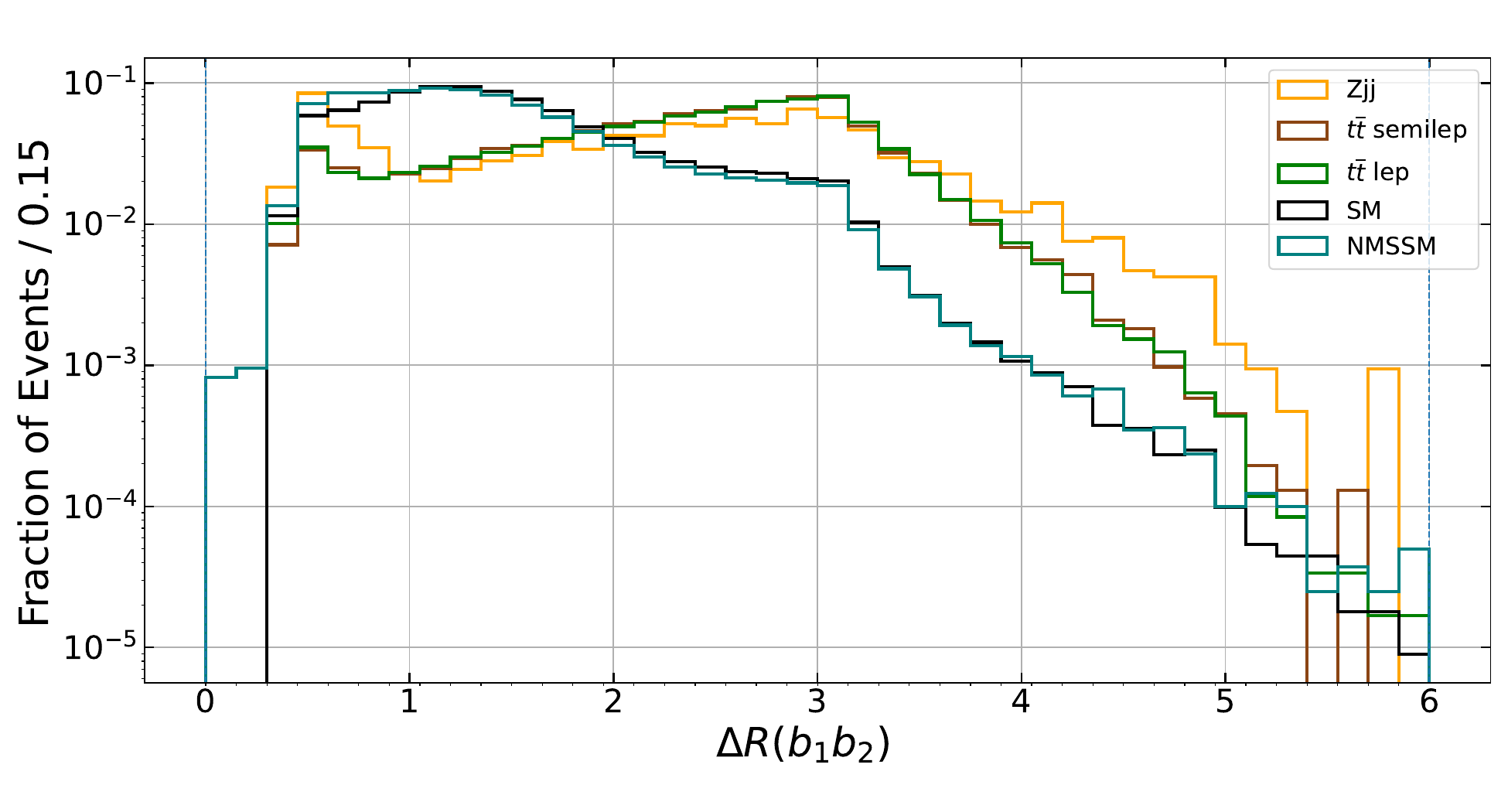}
        \caption{Angular separation $\Delta R(b,b)$.}
    \end{subfigure}
    \hfill
    \begin{subfigure}[b]{0.48\linewidth}
        \centering
        \includegraphics[width=\linewidth, height=6cm]{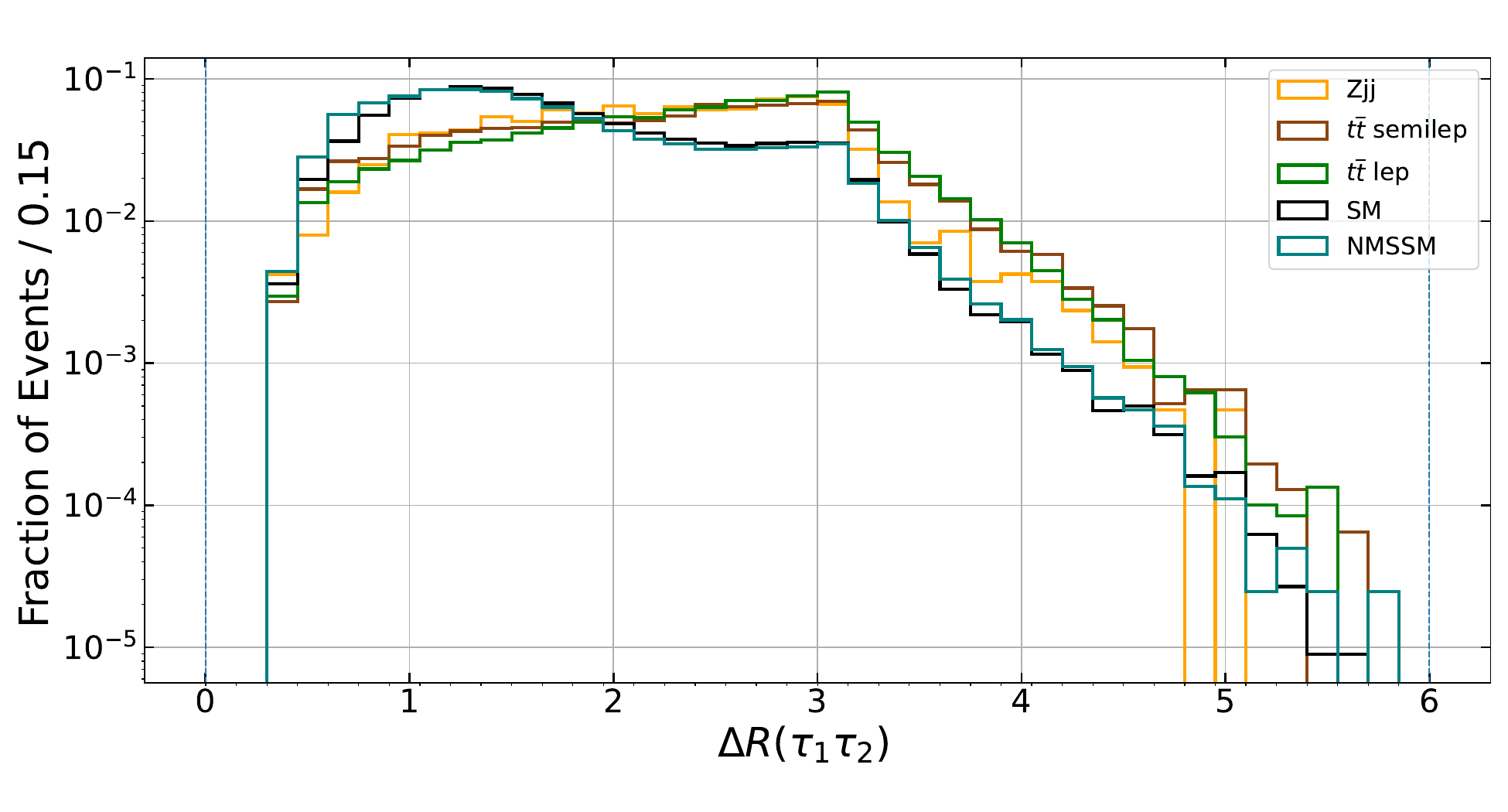}
        \caption{Angular separation $\Delta R(\tau_1,\tau_2)$.}
    \end{subfigure}

    \vspace{0.3cm}

    \begin{subfigure}[b]{0.48\linewidth}
        \centering
        \includegraphics[width=\linewidth, height=6cm]{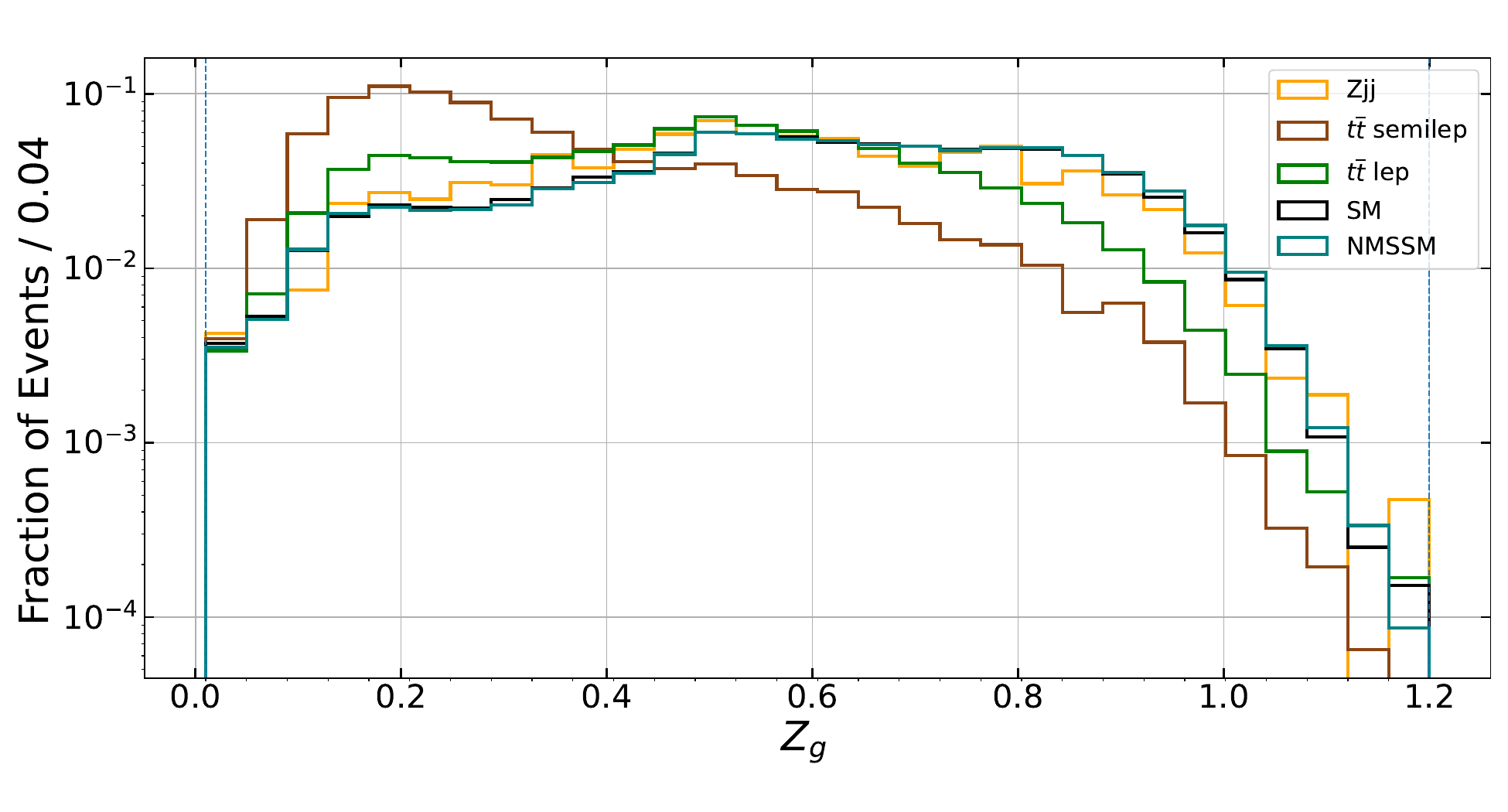}
        \caption{Pion momentum fraction $Z_g$.}
    \end{subfigure}
    \hfill
    \begin{subfigure}[b]{0.48\linewidth}
        \centering
        \includegraphics[width=\linewidth, height=6cm]{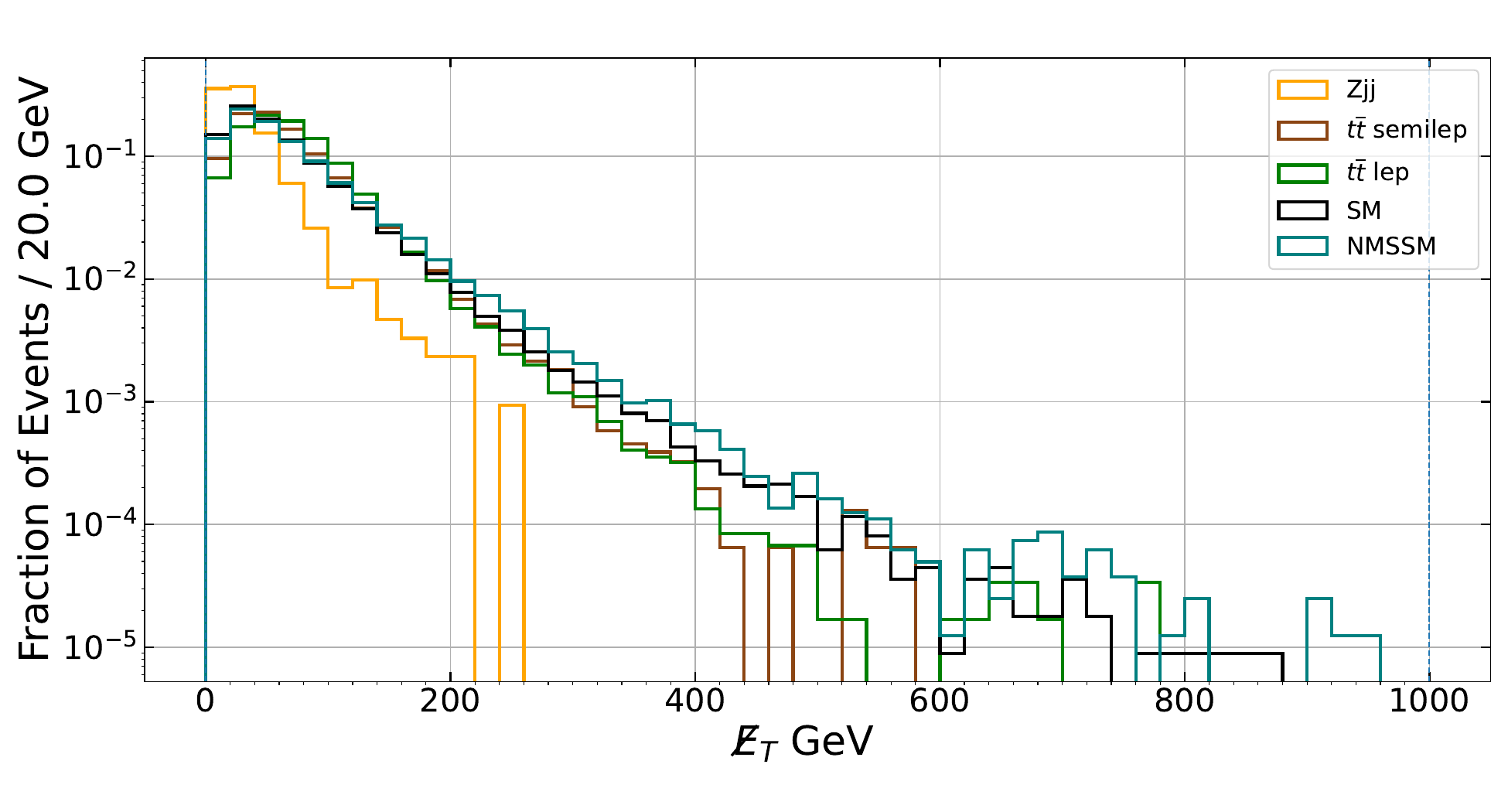}
        \caption{Missing transverse energy $\not\hspace*{-0.085cm} E_T$.}
    \end{subfigure}

    \vspace{0.3cm}

    \begin{subfigure}[b]{0.48\linewidth}
        \centering
        \includegraphics[width=\linewidth, height=6cm]{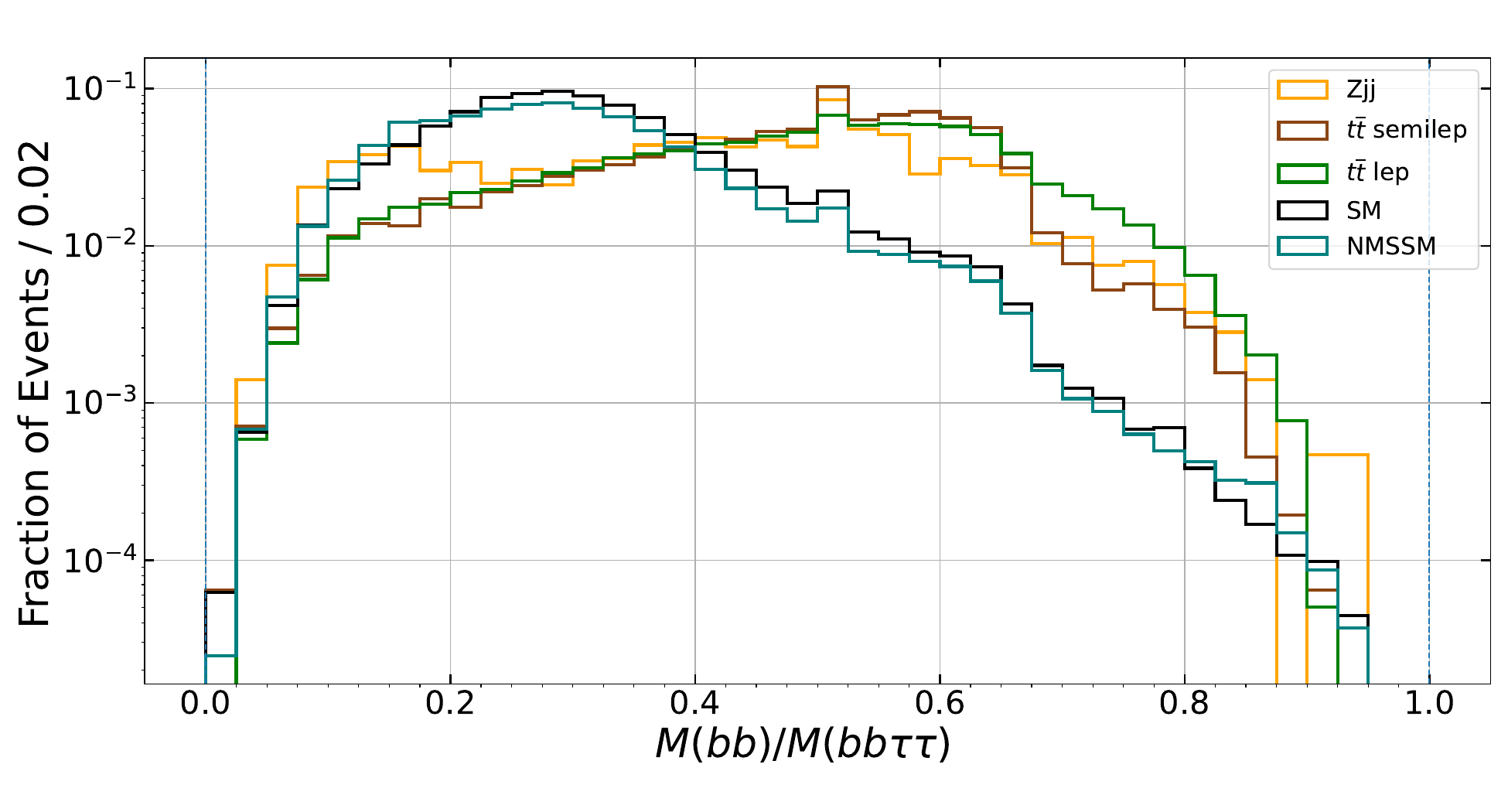}
        \caption{Invariant mass ratio $M(bb)/M(bb\tau\tau)$}
    \end{subfigure}
    \hfill
    \begin{subfigure}[b]{0.48\linewidth}
        \centering
        \includegraphics[width=\linewidth, height=6cm]{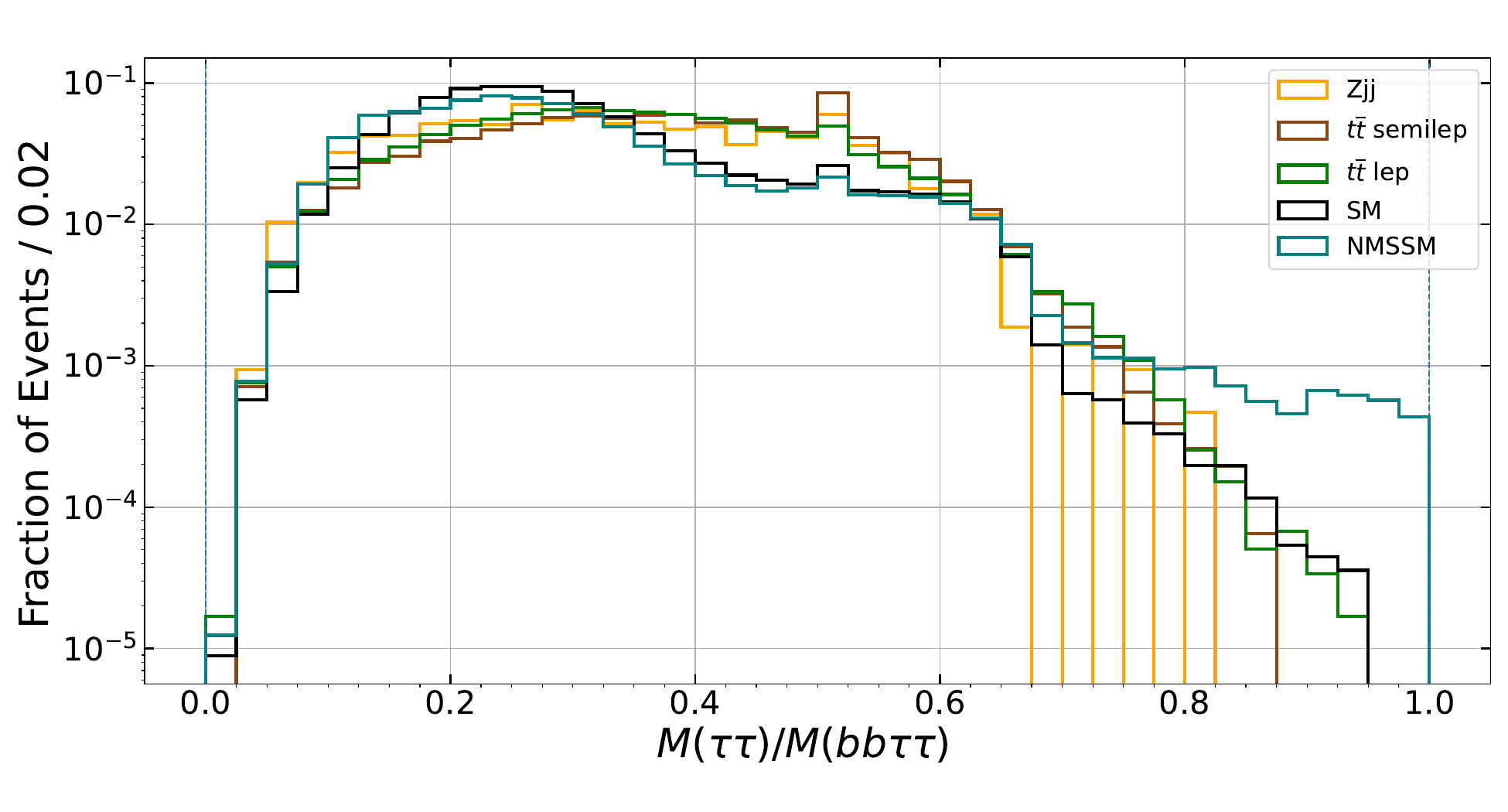}
        \caption{Invariant mass ratio $M(\tau\tau)/M(bb\tau\tau)$}
    \end{subfigure}
    \caption{Normalised histograms after applying Pre-Cut on the NMSSM dataset.}
    \label{fig:NMSSM_plots2}

\end{figure}

The normalised detector-level distributions of the relevant
kinematic observables are presented in
Figs.~\ref{fig:NMSSM_plots1} and
\ref{fig:NMSSM_plots2}. Normalizing the distributions to unity
allows a direct comparison of their shapes independently of the
production cross sections, thereby highlighting the observables that
provide the strongest discrimination between the signal
and the dominant SM backgrounds. These distributions
form the basis for constructing the optimised cut-based analysis
presented below.


The reconstructed invariant masses of the two SM Higgs candidates,
$M(bb)$ and $M(\tau\tau)$, shown in
Figs.~\ref{fig:NMSSM_plots1}(a) and (b), exhibit {moderate enhancements} 
around the Higgs boson mass for both the SM and NMSSM
signals, as expected from the common decay
$h\rightarrow b\bar b$ and
$h\rightarrow\tau^+\tau^-$. Detector resolution, jet-energy
smearing and the presence of neutrinos from hadronic $\tau$ decays
broaden the reconstructed peaks compared to the parton-level
expectation, particularly for the $\tau\tau$ system. Nevertheless,
both signal samples remain significantly more localised around the
Higgs mass than the $t\bar t$ and $Zjj$ backgrounds, whose jet and
lepton combinations arise from unrelated production mechanisms.
These observables therefore provide {good discrimination against
the  backgrounds} (possibly aside from the $t\bar t$ one), although, by construction, they offer
only limited separation between the NMSSM and the SM
di-Higgs signal.

A much stronger distinction between these two contributions is
provided by the reconstructed four-body invariant mass,
$M(bb\tau\tau)$, shown in
Fig.~\ref{fig:NMSSM_plots1}(c). In the SM, Higgs-pair
production proceeds through non-resonant box and triangle diagrams,
resulting in a relatively broad invariant mass spectrum. In contrast,
the NMSSM result receives an additional resonant contribution through the
heavy scalar, i.e., 
$pp\rightarrow H\rightarrow hh$,
which populates the high-mass region and produces substantially more
boosted Higgs bosons. Although detector effects and the invisible
energy carried away by neutrinos smear the resonance peak, the NMSSM
signal still exhibits a noticeably harder $M(bb\tau\tau)$
distribution than the SM process. This observable
therefore provides one of the most powerful handles for identifying
resonant Higgs-pair production and forms the basis of the subsequent
event selection.

The remaining observables in
Fig.~\ref{fig:NMSSM_plots1}(d)--(f) further reflect the more
energetic nature of the resonant NMSSM signal. Since, as mentioned, the SM Higgs bosons are
produced in the decay of a heavy resonance, their decay products
inherit a larger Lorentz boost than in the non-resonant SM process, leading to systematically harder distributions also in the
total transverse hadronic activity, $H_T$, as well as in the
transverse momenta of the leading $b$-jet and leading $\tau$-jet.
The difference between the NMSSM and SM signals is
moderate but clearly visible while both remain considerably harder
than the dominant $t\bar t$ and $Zjj$ backgrounds. Although these
observables alone cannot isolate the resonant contribution, their
combined use significantly enhances the sensitivity of the analysis
when employed together with the resonance reconstruction variables.

The observables shown in
Fig.~\ref{fig:NMSSM_plots2} provide complementary information to the
invariant mass and transverse momentum variables discussed above.
While the latter primarily exploit the boosted kinematics associated
with resonant Higgs production, the observables presented here probe
the geometry of the event and the polarisation properties of the
visible decay products, thereby providing additional discrimination
between the Higgs-pair signal and the dominant SM
backgrounds.

The angular separation variables,
$\Delta R(bb)$ and $\Delta R(\tau\tau)$,
shown in
Figs.~\ref{fig:NMSSM_plots2}(a) and (b),
reflect the different production dynamics of the {heavy} Higgs boson.
Their decay products become more collimated than
those arising from the non-resonant SM process or from
the dominant $t\bar t$ and $Zjj$ backgrounds, resulting in smaller
typical values of $\Delta R$. Although the separation between the
NMSSM and the SM di-Higgs signal is moderate, both
signals occupy noticeably smaller angular separations than the
backgrounds, making these observables useful in the later stages of
the event selection.

A distinctive feature of the present analysis is the inclusion of
the leading pion momentum fraction,
$
Z_g,
$
shown in
Fig.~\ref{fig:NMSSM_plots2}(c).
Unlike conventional kinematic observables, $Z_g$ is sensitive to the
polarisation of the parent $\tau$ lepton through the visible energy
carried by the leading charged pion. In the dominant
$t\bar t$ background, the $\tau$ leptons originate primarily from
the decay
$
W^\pm\rightarrow\tau^\pm\nu,
$
where the charged-current $V-A$ interaction produces predominantly
left-handed $\tau^-$'s (and right-handed $\tau^+$'s). In contrast, the
$\tau$ leptons in both the SM and NMSSM signals are
produced through the decay of the scalar Higgs boson,
$
h\rightarrow\tau^+\tau^-,
$
for which the two $\tau$ leptons are produced with opposite
helicities. This difference in spin correlations modifies the
sharing of the visible momentum between the charged pion and the
associated neutrino, leading to distinct $Z_g$ distributions for the
signal and background processes. Consequently, $Z_g$ provides
information that is largely orthogonal to the purely kinematic
variables discussed above and offers additional discrimination
against the dominant $t\bar t$ background.
The $\not\hspace*{-0.125cm} E_T$ distribution,
shown in
Fig.~\ref{fig:NMSSM_plots2}(d),
is largely governed by the neutrinos produced in the hadronic
$\tau$ decays and, in the case of the $t\bar t$ background, by the
additional neutrinos originating from the leptonic $W$ decays.
Although the signal generally exhibits slightly smaller $\not\hspace*{-0.085cm} E_T$ than the $t\bar t$ background, the separation is
not sufficiently pronounced to motivate a dedicated selection
requirement. We therefore retain this observable primarily as a
consistency check of the detector-level event reconstruction.

Finally, the invariant mass ratios
$M(bb)/M(bb\tau\tau)$ and
$M(\tau\tau)/M(bb\tau\tau)$,
shown in
Figs.~\ref{fig:NMSSM_plots2}(e) and (f),
exploit the strong correlation between the reconstructed SM Higgs
bosons and the parent heavy scalar. Since these Higgs bosons in the
NMSSM signal originate from the decay of a single heavy resonance,
these ratios occupy a relatively narrow region of phase space,
whereas the corresponding background distributions are considerably
broader owing to their non-resonant production mechanisms. The
combination of these correlation observables with the resonance
reconstruction and boosted-event kinematics therefore provides an
efficient strategy for suppressing the remaining backgrounds.

The above distributions clearly demonstrate
that no single observable is sufficient to achieve an efficient
separation between the resonant NMSSM signal and the SM
backgrounds. While the reconstructed Higgs masses provide excellent
rejection of the dominant $t\bar t$ and $Zjj$ processes, they exhibit
only limited discrimination between the NMSSM and the irreducible 
SM di-Higgs production. Conversely, observables such as
$H_T$, $M(bb\tau\tau)$ and the transverse momenta of the Higgs decay
products are considerably more sensitive to the resonant production
mechanism, reflecting the larger Lorentz boost of the Higgs bosons
originating from the heavy scalar decay. Furthermore, the
polarization-sensitive observable $Z_g$ provides an independent
handle for suppressing the remaining $t\bar t$ background by
exploiting the different helicity structure of the $\tau$ leptons.

Guided by these observations, we construct an optimised sequence of
selection criteria aimed at maximizing the statistical sensitivity to
the NMSSM signal. The cuts are applied successively, allowing the effect of each individual
selection on both the signal and the background yields to be
quantified. The final set of optimised cuts adopted in the present
analysis is as follows\footnote{Hereafter, a suffix 1 refers to the leading (i.e., highest $p_T$) object considered.}:
\begin{itemize}
    \item [Pre-Cut:] No. of b-tagged jets, $N_b \geq 2$ and No. of $\tau$-tagged jets, $N_{\tau} \geq 2$.
    \item [Cut A:] $H_T \geq 400 ~{\rm GeV}$
    \item [Cut B:] $M(bb)\ /\ M(bb\tau\tau) \leq 0.3,\ M(\tau\tau)\ /\ M(bb\tau\tau) \leq 0.3$
    \item [Cut C:] $Z_g \geq 0.4$
    \item [Cut D:] $ 90 ~{\rm GeV} \leq M(bb) \leq 150 ~{\rm GeV}$
    \item [Cut E:] $ 95 ~{\rm GeV} \leq M(\tau\tau) \leq 150 ~{\rm GeV}$
    \item [Cut F:] $\Delta R (bb) \leq 1.5$
\end{itemize}

Since the cuts are applied sequentially, the cut-flow provides
valuable information on the relative importance of each selection in
suppressing the dominant backgrounds while retaining the resonant
di-Higgs signal.

\begin{table}[htpb!]
    \centering
    \begin{tabular}{c||c|c|c|c|c}
        \toprule
        \multicolumn{6}{c}{
        $\sqrt{s}=14~{\rm TeV},\quad
        \mathcal{L}=4000~{\rm fb}^{-1}$} \\
        \toprule
        Dataset
        & SM
        & NMSSM
        & $t\bar t$ lep
        & $t\bar t$ semilep
        & $Zjj$ \\
        \midrule\midrule

        $\sigma~[{\rm fb}]$
        & $3.6188$
        & $5.4282$
        & $11493.6$
        & $137921.6$
        & $259895.8$ \\
        \midrule

        Pre-Cut
        & $530$
        & $832$
        & $910630$
        & $2.72\times10^{6}$
        & $737411$ \\
        \hline

        Cut A
        & $161$
        & $289$
        & $88960$
        & $335976$
        & $44354$ \\
        \hline

        Cut B
        & $84$
        & $151$
        & $1117$
        & $7171$
        & $6236$ \\
        \hline

        Cut C
        & $64$
        & $117$
        & $534$
        & $2022$
        & $5891$ \\
        \hline

        Cut D
        & $55$
        & $96$
        & $198$
        & $1286$
        & $2425$ \\
        \hline

        Cut E
        & $29$
        & $41$
        & $138$
        & $0$
        & $0$ \\
        \bottomrule
    \end{tabular}
    \caption{Cut-flow for the SM di-Higgs process, the NMSSM
    signal, and the dominant backgrounds at the HL-LHC with
    $\sqrt{s}=14~\TeV$ and $\mathcal{L}=4000~{\rm fb}^{-1}$.
    Each cut is applied sequentially, and the corresponding
    surviving event yields, $N_L$, are reported. Higher-order QCD
    corrections are included through $K$-factors of
    $K_{hh}=2.2$, $K_{t\bar t}=1.6$, and $K_{Zjj}=1.2$.
    The quoted cross sections and event yields include these
    corrections.}
    \label{tab:NMSSM_cut_chart}
\end{table}

Several noteworthy features emerge from the cut-flow presented in
Tab.~\ref{tab:NMSSM_cut_chart}. The initial event selection removes
the overwhelming QCD contribution together with a substantial
fraction of the $t\bar t$ and $Zjj$ backgrounds whilst preserving a
large fraction of the signal, indeed, reflecting the more energetic topology
of resonant di-Higgs production. The invariant mass and Higgs
reconstruction requirements provide the largest improvement in the
signal-to-background ratio by selecting events compatible with the
decay of a heavy resonance. The $\tau$-polarisation observable,
$Z_g$, gives an additional reduction of the remaining $t\bar t$
background owing to the different helicity structures of $\tau$
leptons originating from Higgs and top-quark decays.

After the full event selection, the semileptonic $t\bar t$, $Zjj$ and
QCD backgrounds are completely removed within the simulated sample.
However, the dileptonic $t\bar t$ background cannot be eliminated and
remains with a large number of surviving events, exceeding both the SM and NMSSM
 signal yields, yet some sensitivity exists. 

\begin{table}[H]
    \centering
    \begin{tabular}{c||c|c|c|c|c|c}
        \toprule
        Cut
        & $\mathcal{S}_{\rm SM}$
        & $\mathcal{S}^{\rm combined}_{\rm SM}$
        & $\mathcal{S}_{\rm NMSSM}$
        & $\mathcal{S}^{\rm combined}_{\rm NMSSM}$
        & $\mathcal{S}_{\rm signal}$
        & $\mathcal{S}^{\rm combined}_{\rm signal}$ \\
        \midrule

        Pre-Cut
        & $0.254$ & $0.359$
        & $0.398$ & $0.563$
        & $0.144$ & $0.204$ \\

        Cut A
        & $0.235$ & $0.332$
        & $0.422$ & $0.597$
        & $0.187$ & $0.264$ \\

        Cut B
        & $0.696$ & $0.985$
        & $1.251$ & $1.769$
        & $0.554$ & $0.783$ \\

        Cut C
        & $0.695$ & $0.984$
        & $1.270$ & $1.796$
        & $0.574$ & $0.812$ \\

        Cut D
        & $0.878$ & $1.241$
        & $1.529$ & $2.163$
        & $0.650$ & $0.919$ \\

        Cut E
        & $2.389$ & $3.379$
        & $3.336$ & $4.717$
        & $0.918$ & $1.298$ \\

        \bottomrule
    \end{tabular}
    \caption{Evolution of the Asimov statistical significance,
    $\mathcal{S}_{\rm A}$, for the SM and NMSSM di-Higgs
    processes after each stage of the event selection at the
    HL-LHC with $\sqrt{s}=14~\TeV$ and
    $\mathcal{L}=4000~{\rm fb}^{-1}$. For the individual SM
    and NMSSM significances, the corresponding signal yield is
    evaluated against the sum of the $t\bar t$ dileptonic,
    $t\bar t$ semileptonic, and $Zjj$ background yields.
    The quantity $\mathcal{S}_{\rm signal}$ characterises the
    separation between the SM and NMSSM hypotheses, taking
    $S=|N_{\rm NMSSM}-N_{\rm SM}|$ and
    $B=N_{\rm SM}+N_{\rm bkg}$. The ATLAS and CMS datasets are
    assumed to be statistically independent and to have equal
    sensitivities, giving
    $\mathcal{S}^{\rm combined}=\sqrt{2}\,\mathcal{S}_{\rm A}$.}
    \label{tab:NMSSM_significance}
\end{table}

The corresponding significances are presented in
Tab.~\ref{tab:NMSSM_significance}. Here, we use the formula mentioned in Eq.~(\ref{eq:Significance}). The optimisation procedure
steadily improves the sensitivity, yielding final significances for the genuine NMSSM effects (last two columns) up to $~1.3\sigma$.
In fact, notice that in the table we have also added the corresponding columns for the predicted combined significance across the two multi-purpose HL-LHC experiments (ATLAS and CMS).

{In a more realistic experimental context, however, a further limiting factor is represented by the systematic uncertainties of the background. These cannot be easily predicted in a phenomenological analysis, but they can still be parametrised to evaluate their impact on the limits. For this we introduce the background uncertainties in the Asimov formula, which becomes:
\begin{equation}
Z=\sqrt{2\left[(S+B)\log\left(\frac{(S+B)(B+\Delta B^2)}{B^2+(S+B)\Delta B^2}\right)\right]-\frac{B^2}{\Delta B^2}\log\left(1+\frac{S\ \Delta B^2}{B(B+\Delta B^2)}\right)}\;.
\label{eq:AsimovDB}
\end{equation}
Considering only the $\mathcal{S}^{\rm combined}_{\rm signal}$ values of Tab.~\ref{tab:NMSSM_significance}, the dependence on the systematics is represented in Fig.~\ref{fig:NMSSMZDB} after each selection. After the complete set of cuts is applied, the significance saturates to $Z\simeq1.3$ if $\Delta B\lesssim 1\%$ while for $\Delta B=5\%$ or $\Delta B=10\%$ it drops to $Z_{5\%}=1.1$ and $Z_{10\%}=0.8$, respectively.
\begin{figure}[h!]
\centering
\includegraphics[width=0.5\textwidth]{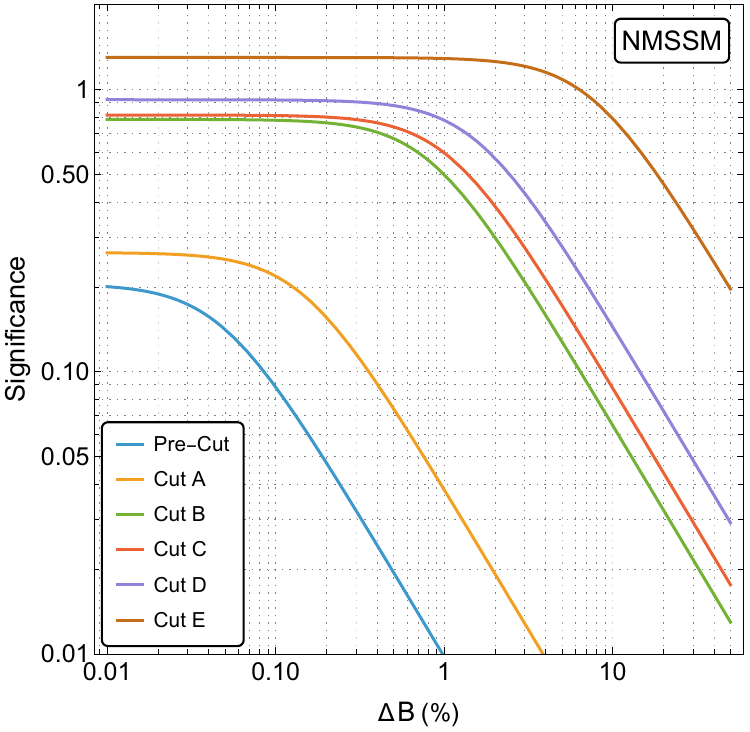}
\caption{\label{fig:NMSSMZDB} Parametric dependence of $\mathcal{S}^{\rm combined}_{\rm signal}$ of the NMSSM scenario on the background systematic uncertainty.}
\end{figure}
}

The analysis demonstrates that the proposed cut strategy is
highly effective in suppressing the dominant reducible backgrounds.
Nevertheless, the final sensitivity is limited by the residual
dileptonic $t\bar t$ background together with the irreducible background due to 
SM di-Higgs contribution {(and by their associated uncertainties)}. Consequently, any further
improvement in the discovery potential will require an improved treatment of the observables that
directly probe the resonant production mechanism, such as the
reconstructed heavy Higgs mass and
$\tau$ polarisation information, or, alternatively, to leverage the 
boosted-event kinematics of the NMSSM component via  
more sophisticated
multi-variate analysis techniques.



\subsection{Detector-level Analysis of the C2HDM BP}
\label{cut_C2HDM}

We now perform the same detector-level analysis for the C2HDM
BP introduced in Sec.~\ref{sec:C2HDM}. The event generation,
detector simulation and object reconstruction follow exactly the
same procedure as described for the NMSSM BP, allowing a
direct comparison between the two scenarios. Consequently, any
differences observed in the detector-level distributions originate
from the underlying production dynamics of the two models rather
than from differences in the analysis strategy.

Similarly to the NMSSM, where the dynamics of resonant SM Higgs-pair
production is governed by both the extended Higgs and top-quark sector (i.e., stop companions), the
C2HDM also receives contributions from the CHM dynamics.
In particular, heavy scalar production proceeds through loops
containing not only the SM top quark but also the eight
HTPs discussed in Sec.~\ref{sec:C2HDM}. The
presence of the additional $T_iT_jH$, $T_iT_jh$ and $T_iT_ihh$
interactions modifies both the production rate and the kinematic
properties of the Higgs-pair system. In particular, given that current experimental limits on the HTP masses are stronger than those on stop masses and that $m_H$ values are generally higher in the C2HDM than in the NMSSM \cite{DeCurtis:2018iqd} (indeed, our BP choices in Tabs.~\ref{tab:NMSSM_benchmark}--\ref{tab:C2HDM_benchmark} reflect this),
a consequence of this is that the resonant
C2HDM signal is expected to exhibit an even harder 
kinematic spectrum than the NMSSM signal.

\begin{figure}[htpb!]
    \centering

    \begin{subfigure}[b]{0.48\linewidth}
        \centering
        \includegraphics[width=\linewidth, height=6cm]{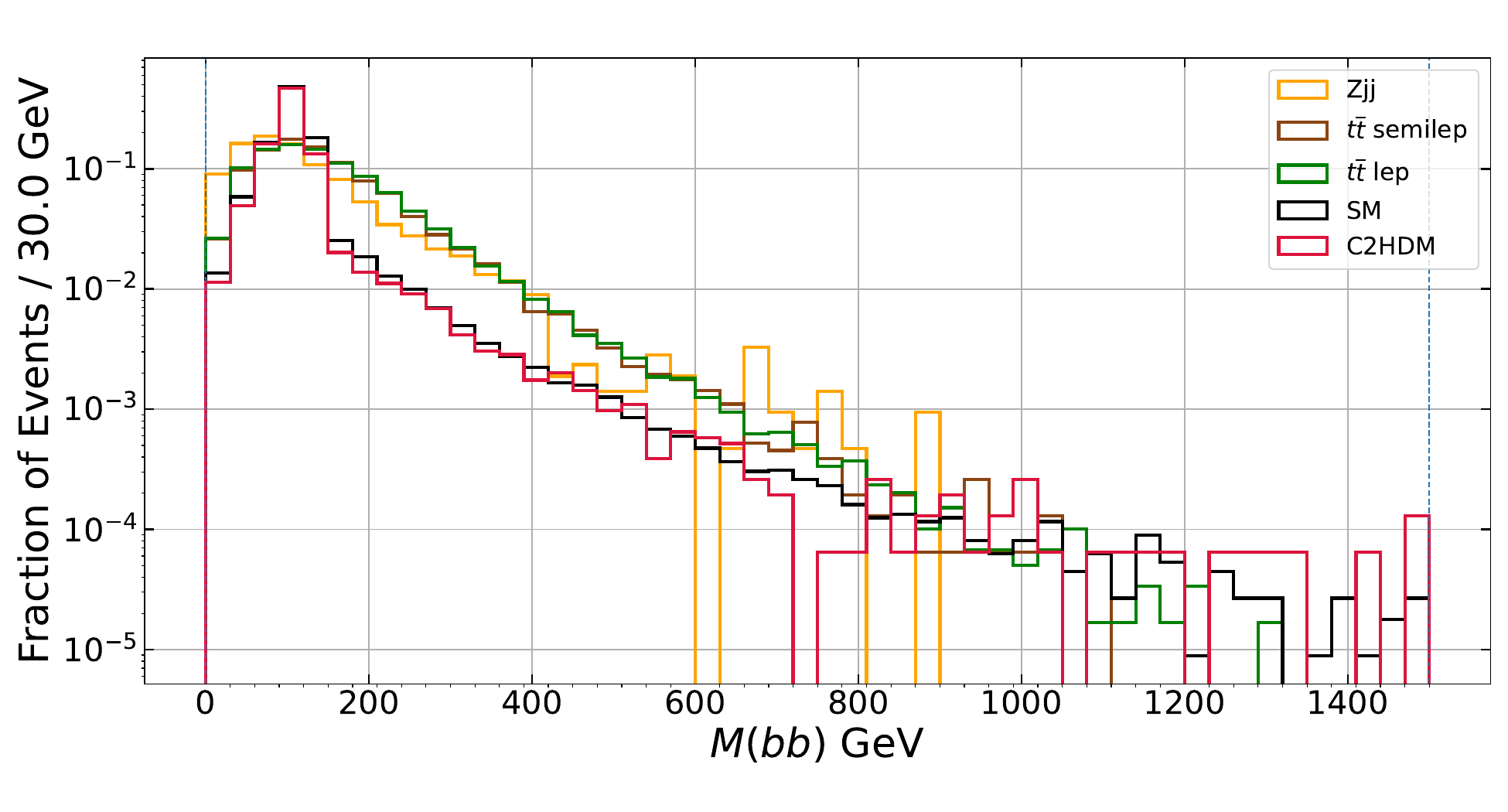}
        \caption{Invariant mass $M(bb)$.}
    \end{subfigure}
    \hfill
    \begin{subfigure}[b]{0.48\linewidth}
        \centering
        \includegraphics[width=\linewidth, height=6cm]{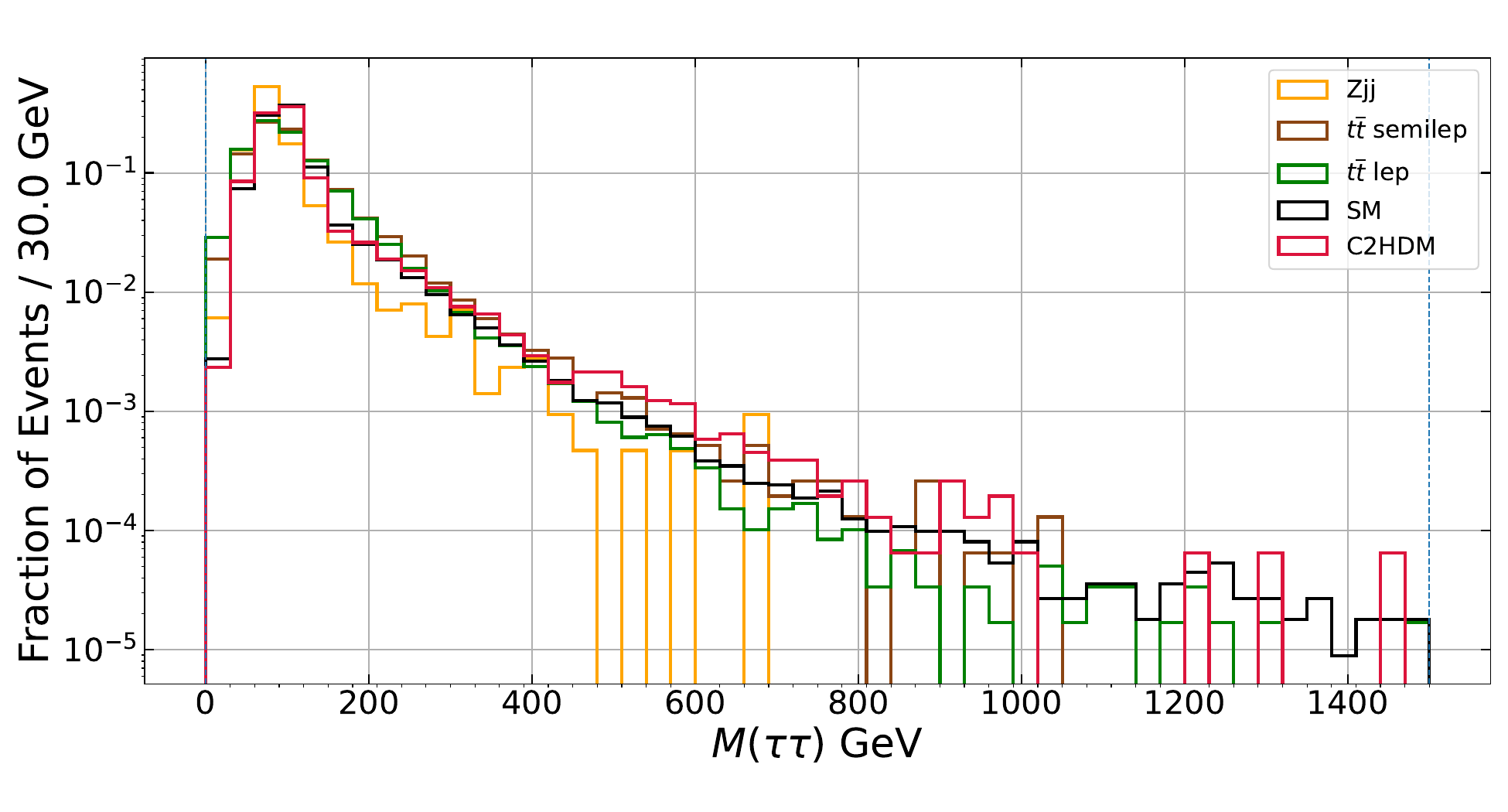}
        \caption{Invariant mass $M(\tau\tau)$.}
    \end{subfigure}

    \vspace{0.3cm}

    \begin{subfigure}[b]{0.48\linewidth}
        \centering
        \includegraphics[width=\linewidth, height=6cm]{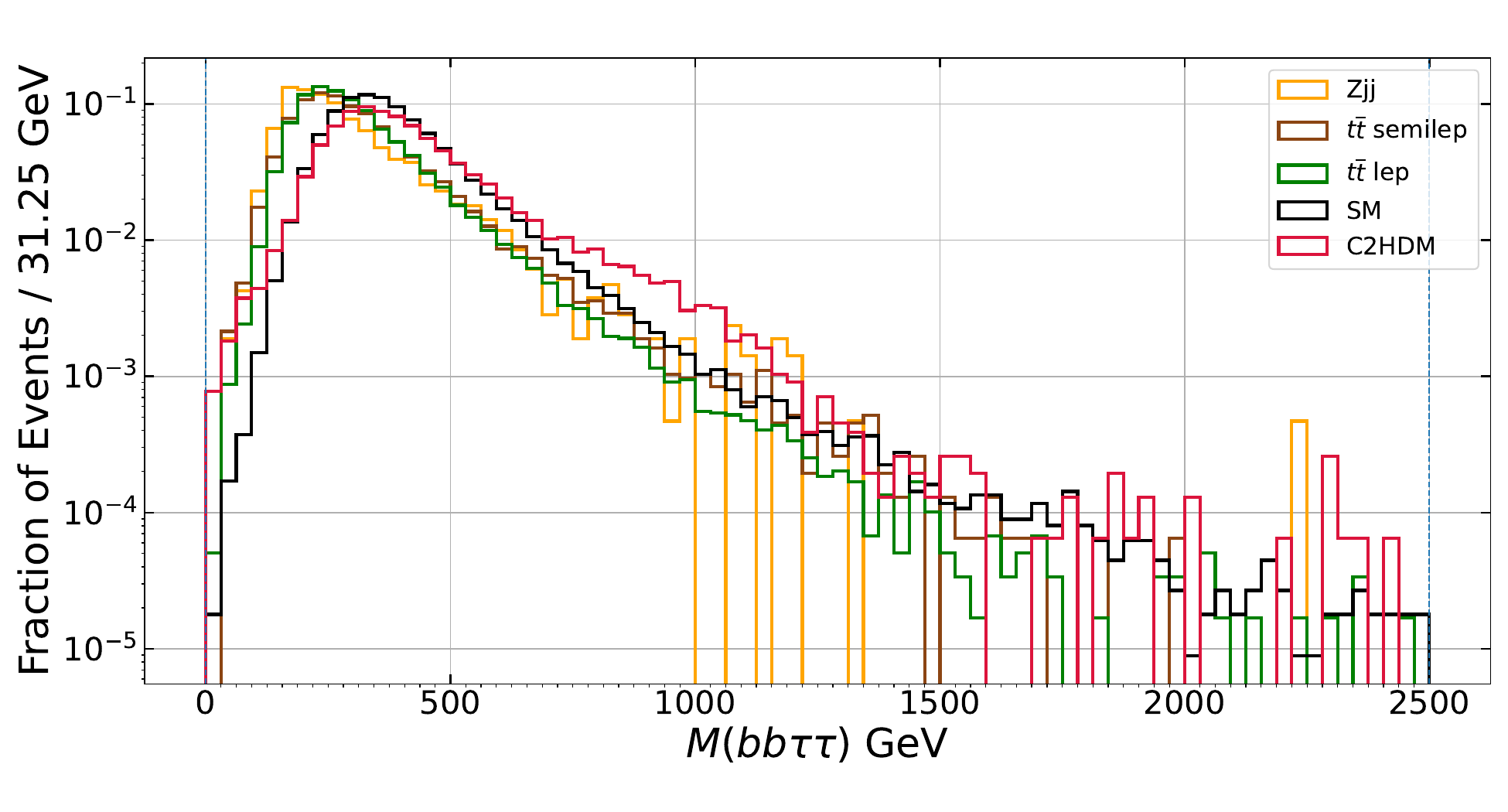}
        \caption{Invariant mass $M(bb\tau\tau)$.}
    \end{subfigure}
    \hfill
    \begin{subfigure}[b]{0.48\linewidth}
        \centering
        \includegraphics[width=\linewidth, height=6cm]{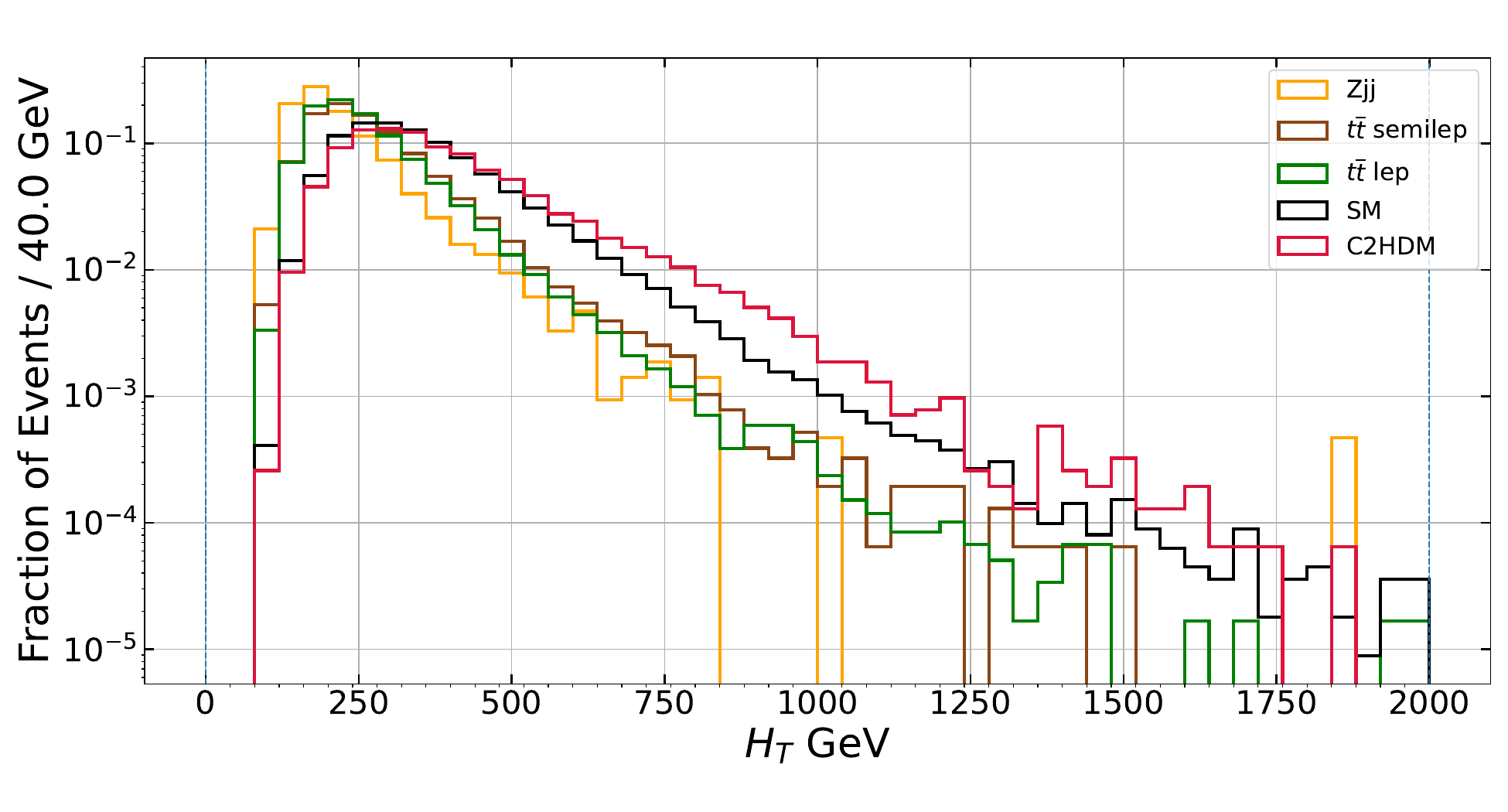}
        \caption{Transverse hadronic activity $H_T$.}
    \end{subfigure}

    \vspace{0.3cm}

    \begin{subfigure}[b]{0.48\linewidth}
        \centering
        \includegraphics[width=\linewidth, height=6cm]{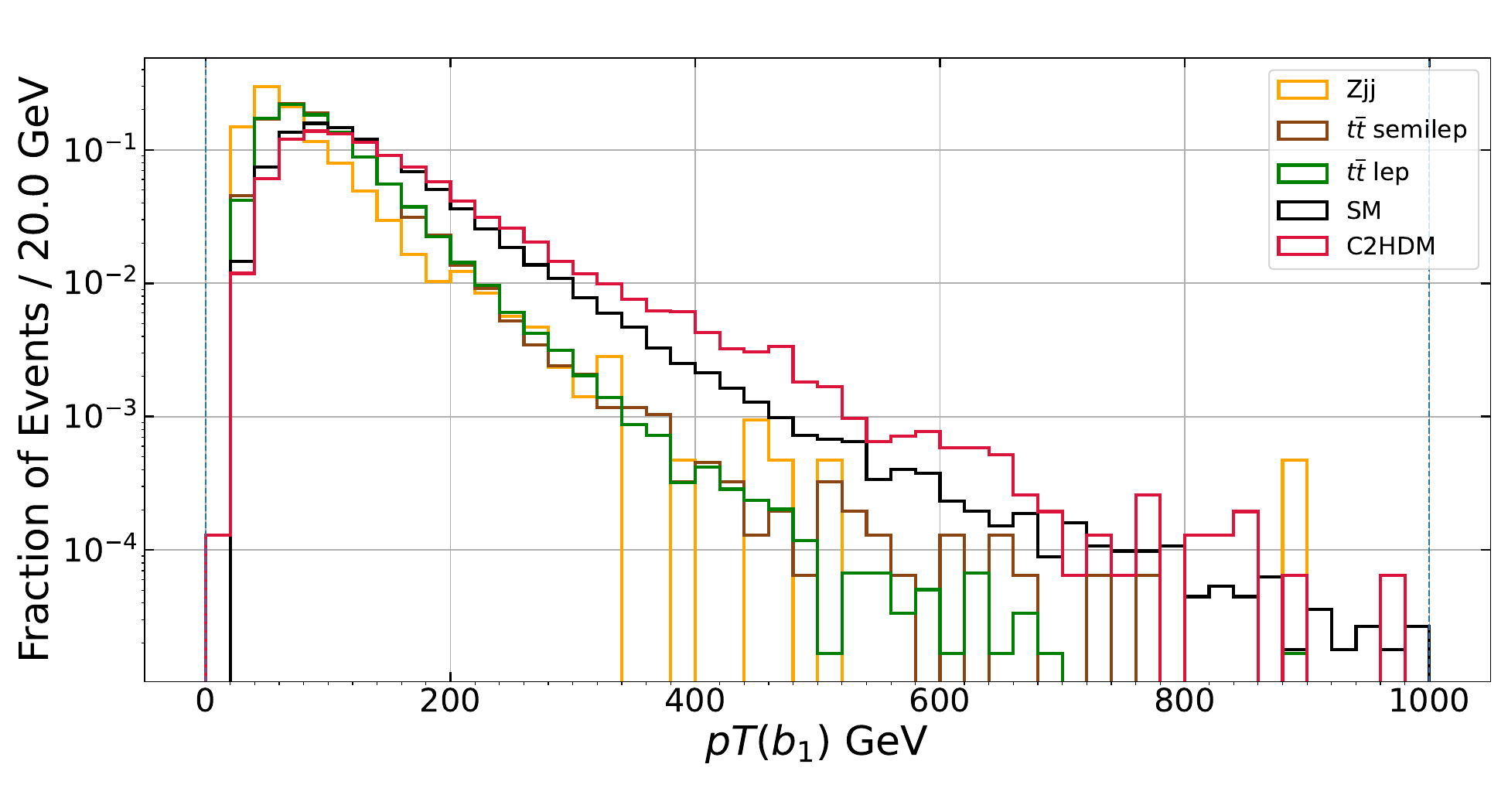}
        \caption{Transverse momentum $p_T(b_1)$.}
    \end{subfigure}
    \hfill
    \begin{subfigure}[b]{0.48\linewidth}
        \centering
        \includegraphics[width=\linewidth, height=6cm]{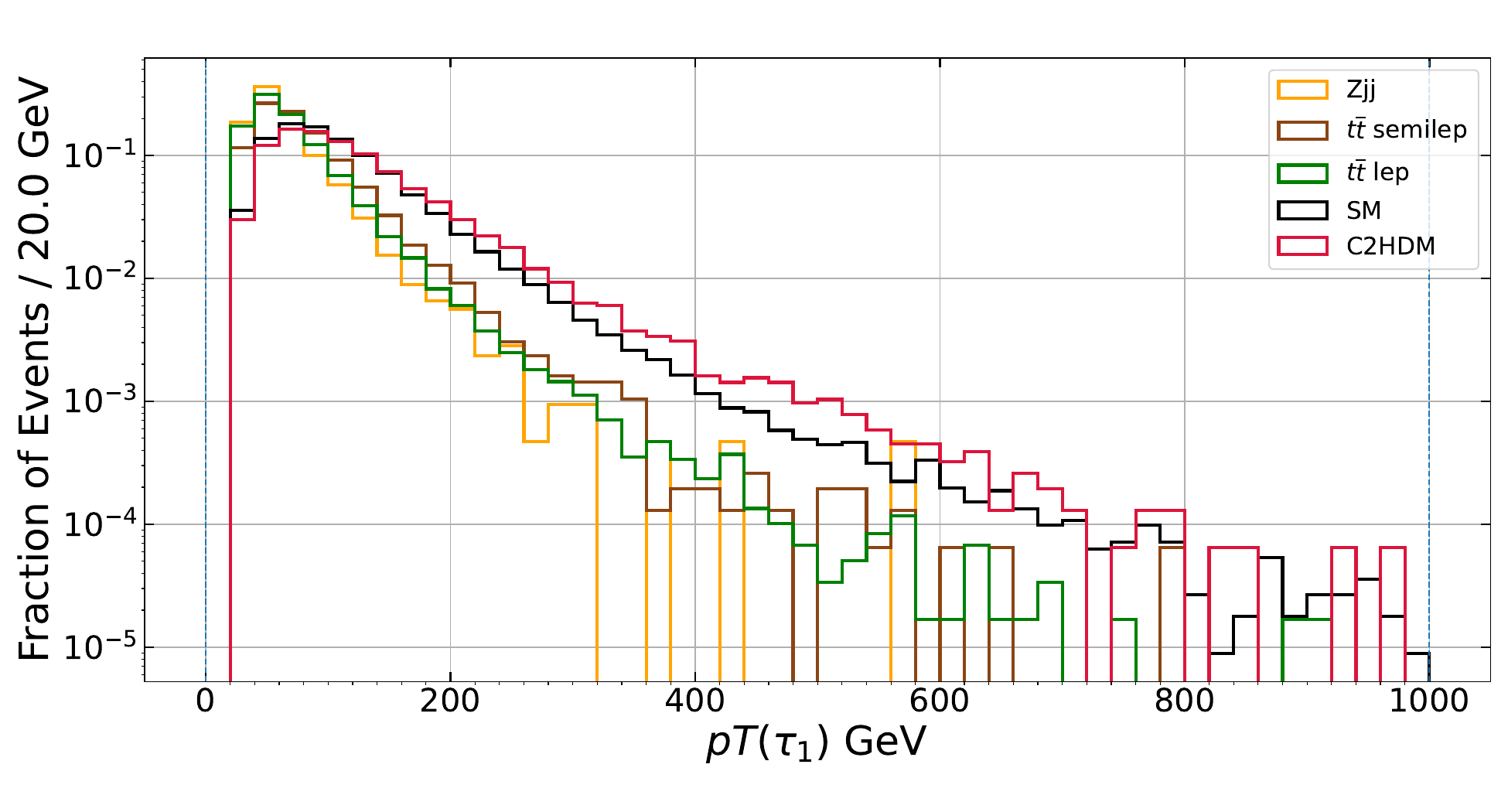}
        \caption{Transverse momentum $p_T(\tau_1)$.}
    \end{subfigure}
    \caption{Normalised distribution of reconstructed final state observables for both signals and major SM backgrounds after Pre-Cut at HL-LHC with $\sqrt{s}$ = 14 TeV for C2HDM dataset.}
    \label{fig:C2HDM_plots1}

\end{figure}

\begin{figure}[htbp]
    \centering

    \begin{subfigure}[b]{0.48\linewidth}
        \centering
        \includegraphics[width=\linewidth, height=6cm]{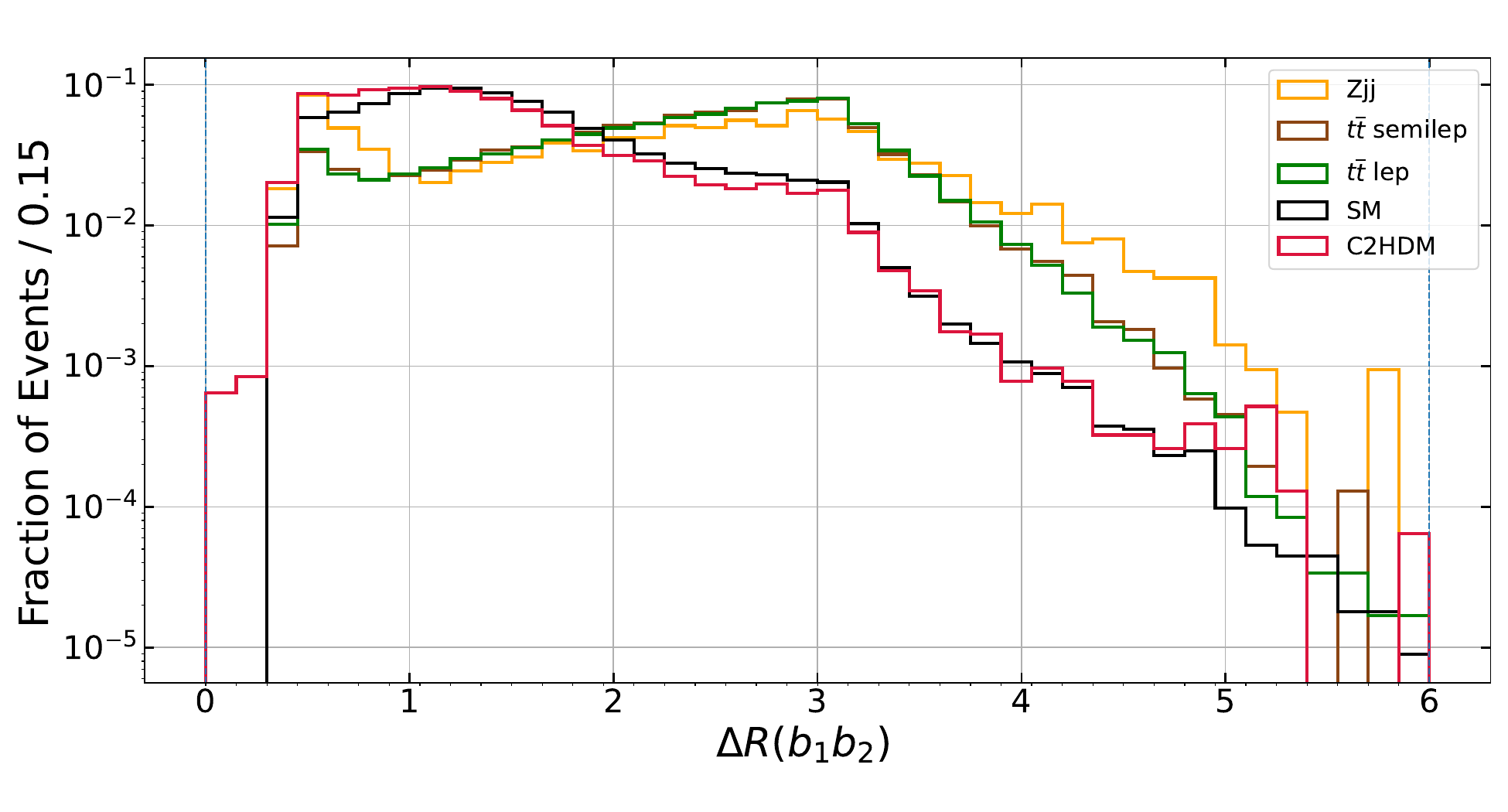}
        \caption{Angular separation $\Delta R(b,b)$.}
    \end{subfigure}
    \hfill
    \begin{subfigure}[b]{0.48\linewidth}
        \centering
        \includegraphics[width=\linewidth, height=6cm]{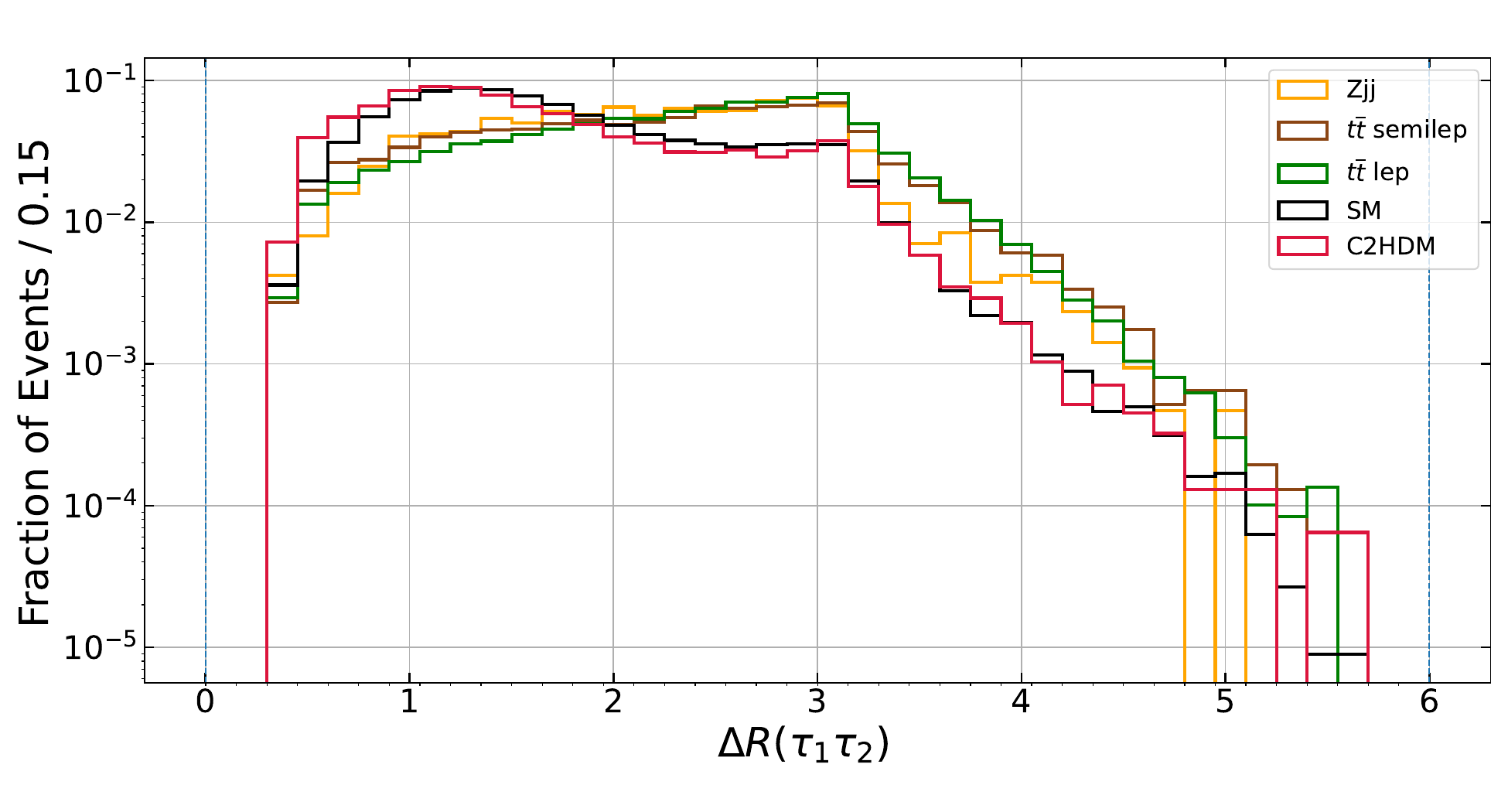}
        \caption{Angular separation  $\Delta R(\tau_1,\tau_2)$.}
    \end{subfigure}

    \vspace{0.3cm}

    \begin{subfigure}[b]{0.48\linewidth}
        \centering
        \includegraphics[width=\linewidth, height=6cm]{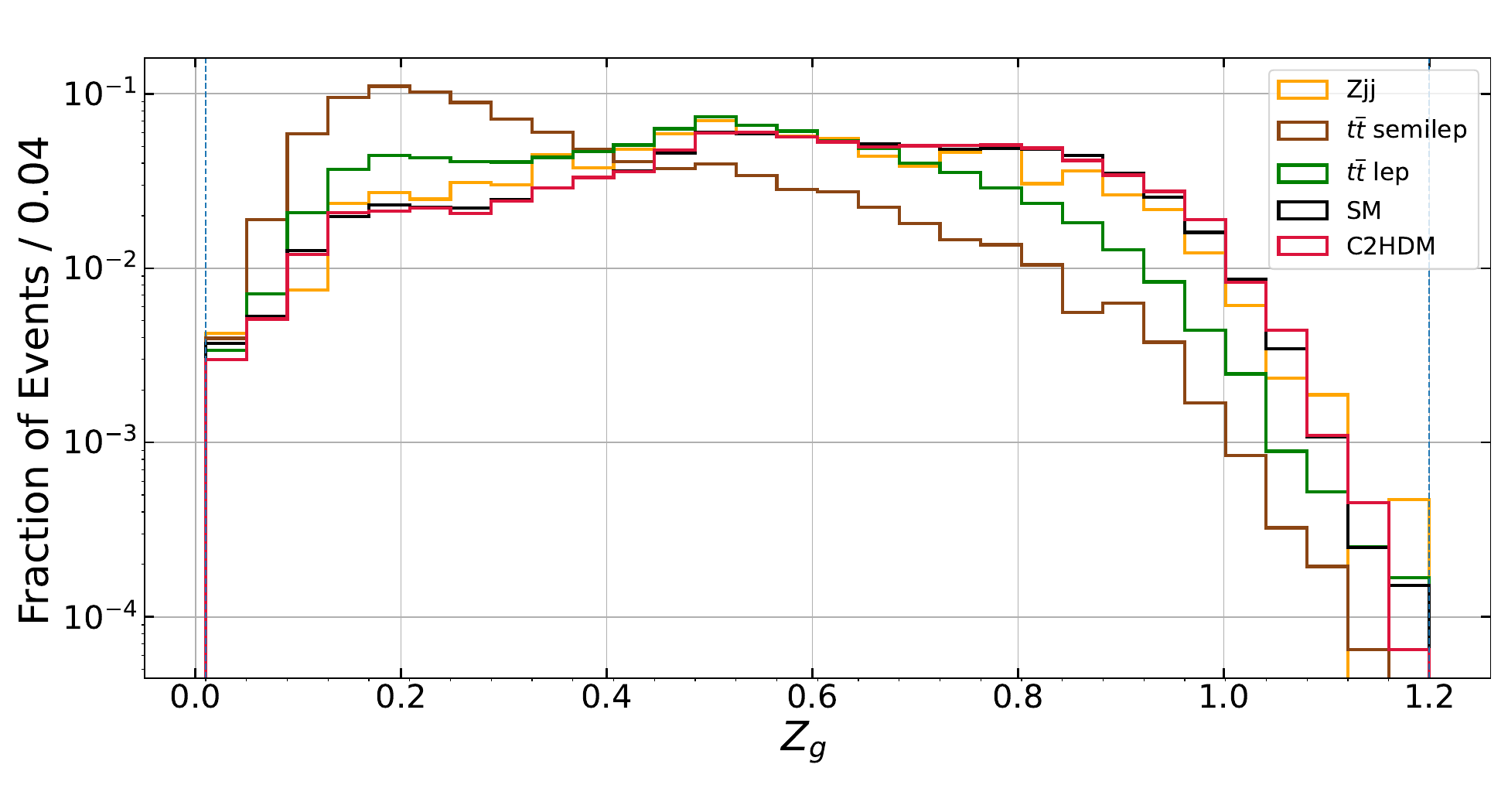}
        \caption{Pion momentum fraction $Z_g$.}
    \end{subfigure}
    \hfill
    \begin{subfigure}[b]{0.48\linewidth}
        \centering
        \includegraphics[width=\linewidth, height=6cm]{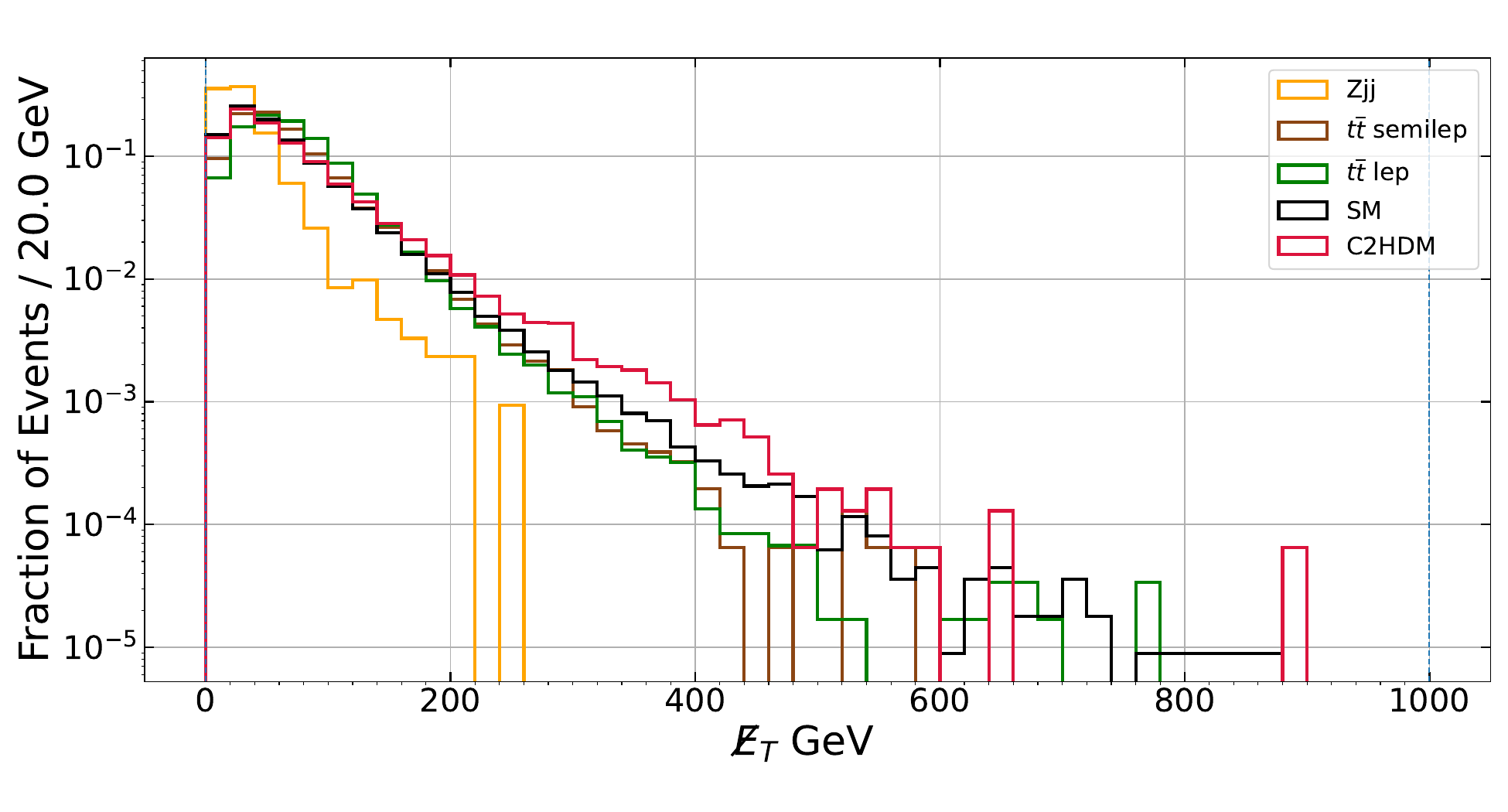}
        \caption{Missing transverse energy $\not\hspace*{-0.085cm}E_T$.}
    \end{subfigure}

    \vspace{0.3cm}

    \begin{subfigure}[b]{0.48\linewidth}
        \centering
        \includegraphics[width=\linewidth, height=6cm]{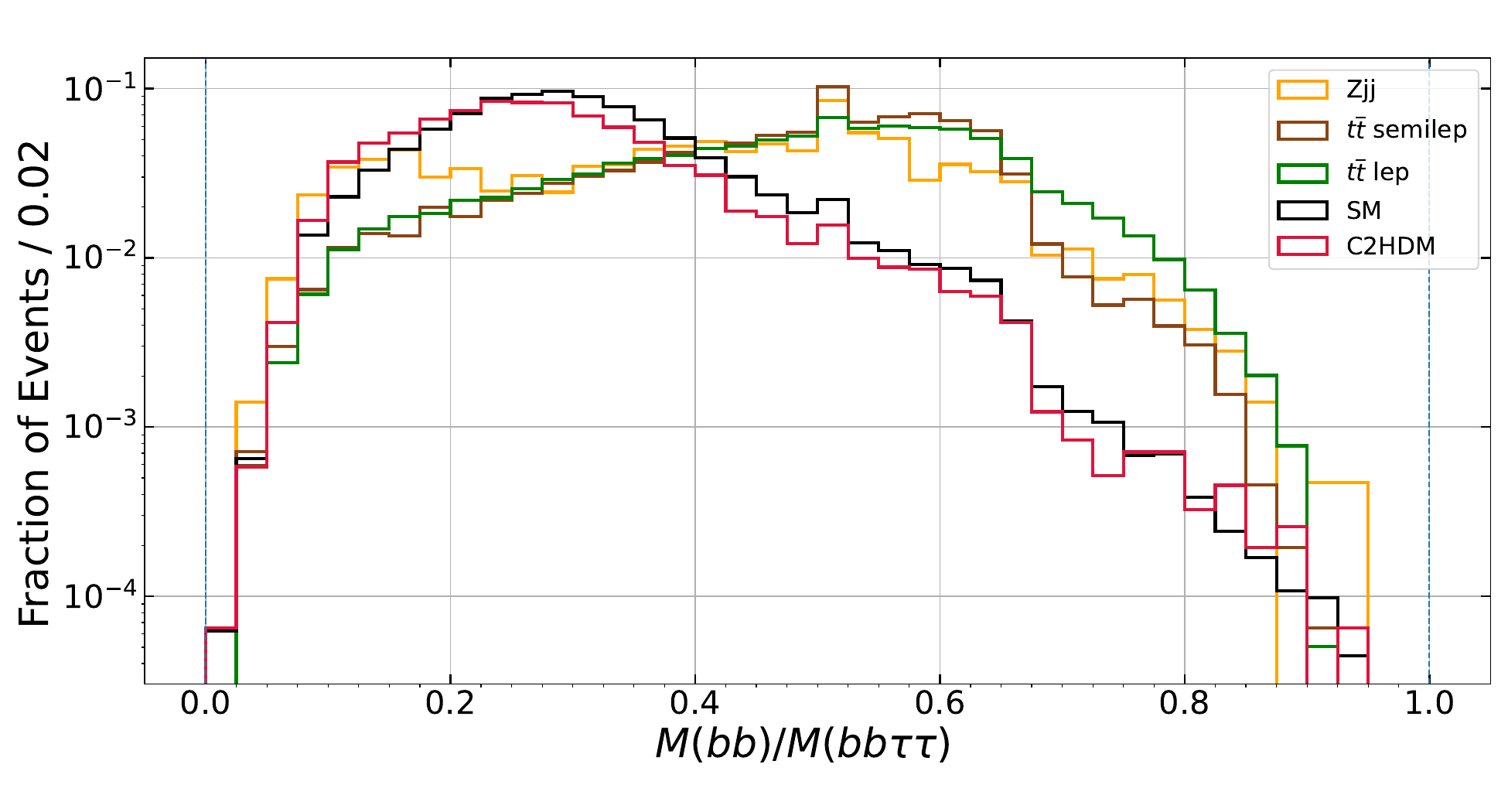}
        \caption{Invariant mass ratio $M(bb)/M(bb\tau\tau)$}
    \end{subfigure}
    \hfill
    \begin{subfigure}[b]{0.48\linewidth}
        \centering
        \includegraphics[width=\linewidth, height=6cm]{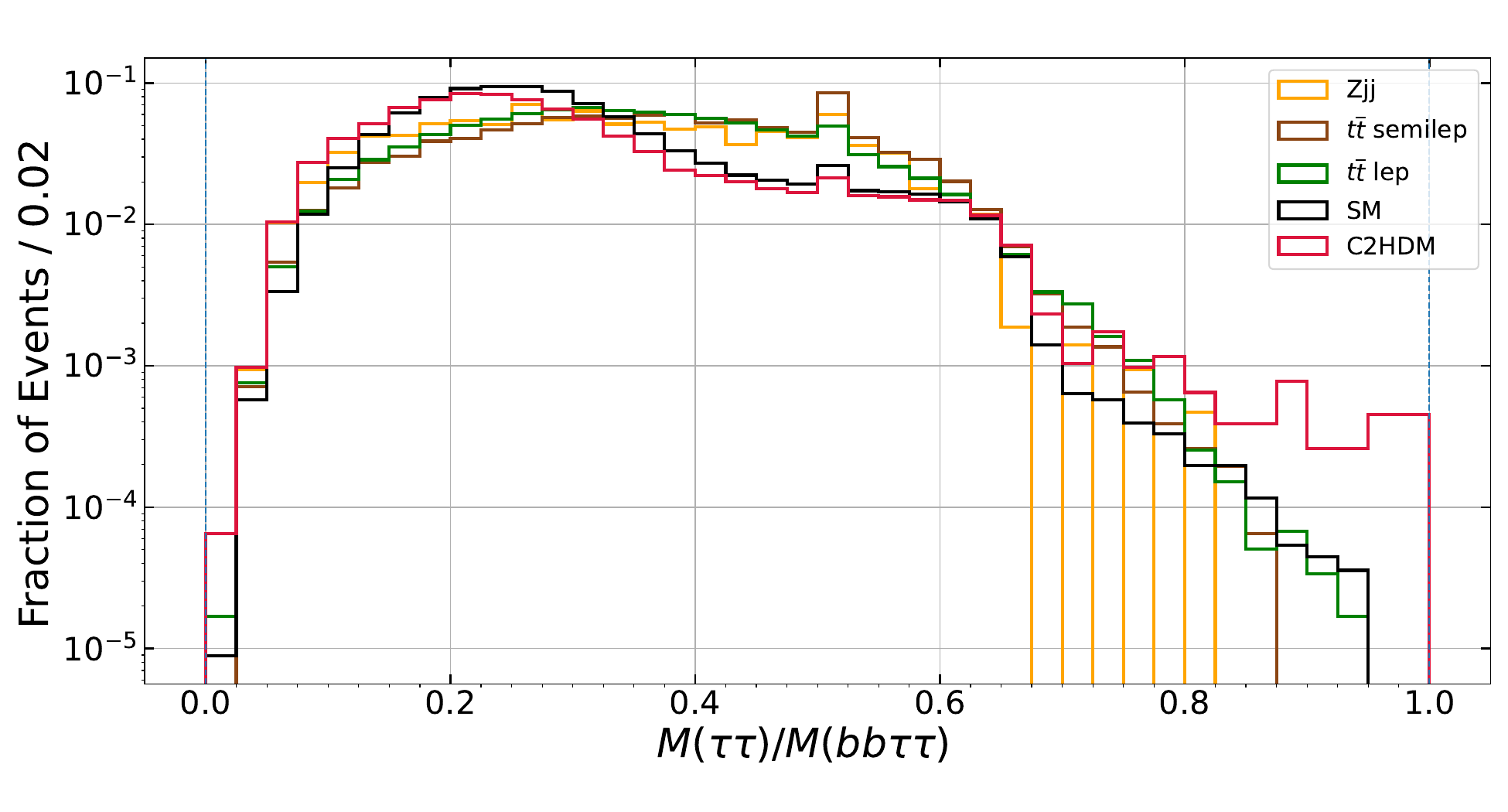}
        \caption{Invariant mass ratio $M(\tau\tau)/M(bb\tau\tau)$}
    \end{subfigure}
    \caption{Histograms after applying Pre-Cut on the C2HDM dataset.}
    \label{fig:C2HDM_plots2}

\end{figure}

The normalised detector-level distributions of the relevant
kinematic observables are presented in
Figs.~\ref{fig:C2HDM_plots1} and
\ref{fig:C2HDM_plots2}. Besides providing discrimination against the
dominant SM backgrounds, these distributions allow us to
investigate how the composite nature of the Higgs sector manifests
itself in detector-level observables. As intimated and as it will be shown below,
the larger Lorentz boost of the Higgs bosons, together with the
modified production mechanism, induced by the heavy scalar resonance
and the HTP loops, leads to a noticeably improved
separation between the C2HDM signal and the SM 
backgrounds (both irreducible and irreducible ones). These features ultimately translate into a more
efficient cut-based analysis than that obtained for the NMSSM
case.

The detector-level distributions shown in
Fig.~\ref{fig:C2HDM_plots1} exhibit the characteristic features of
resonant Higgs-pair production in the C2HDM. While many of the observables display behaviours similar to
that observed in the NMSSM analysis, harder kinematic spectra are also visible, as expected.
Furthermore, 
the reconstructed SM Higgs candidate masses,
$M(bb)$ and $M(\tau\tau)$,
shown in
Figs.~\ref{fig:C2HDM_plots1}(a) and (b),
continue to provide an efficient selection. Since both the SM and C2HDM signals contain the
same decays,
$
h\rightarrow b\bar b,\qquad
h\rightarrow\tau^+\tau^-,
$
their reconstructed mass distributions peak around the Higgs boson
mass and remain broadly visible after detector simulation.
Consequently, these observables are again useful for primarily rejecting
the dominant $t\bar t$ and $Zjj$ backgrounds while providing only
limited discrimination between the resonant C2HDM contribution and
the irreducible background due to the SM di-Higgs process.

A much clearer distinction between the two di-Higgs processes (and backgrounds) is observed in the reconstructed
four-body invariant mass,
$M(bb\tau\tau)$,
shown in
Fig.~\ref{fig:C2HDM_plots1}(c).
Unlike the SM process, where Higgs-pair production is
dominated by non-resonant box and triangle diagrams, as expected, the C2HDM
contains an additional (rather) heavy scalar resonance together with modified
ggF amplitudes arising from the HTP sector.
Although detector resolution and the $\not\hspace*{-0.085cm} E_T$ carried by the
neutrinos from the $\tau$ decays smear the resonance peak, the
signal retains a pronounced enhancement in the high-mass region.
This observable therefore provides again the strongest sensitivity to the
 resonant production mechanism and constitutes one of the
most powerful discriminants used in the subsequent event selection.

The described boosted nature of the resonant signal (limitedly to the C2HDM component) is further reflected in
the distributions of the total transverse hadronic activity,
$H_T$, and the transverse momenta of the leading $b$-jet and leading
$\tau$-jet shown in
Figs.~\ref{fig:C2HDM_plots1}(d)--(f).
Compared with both SM di-Higgs production and the
NMSSM case, the C2HDM signal exhibits systematically harder
spectra here too, reflecting the larger momentum transferred to the Higgs
bosons during the decay of the heavy scalar. Furthermore, here, the interference
between the SM top-quark loop and the additional
HTP loops modifies not only the total production rate (like in the NMSSM, albeit in opposite direction)
but also the momentum spectrum of the heavy scalar. Consequently, this also contributes to a 
larger fraction of SM Higgs bosons produced with sizeable
Lorentz boosts, leading to the harder $H_T$, $p_T(b_1)$ and
$p_T(\tau_1)$ distributions observed in the detector-level
simulation. These observables, therefore, provide significantly
improved discrimination not only against the dominant $t\bar t$ and
$Zjj$ backgrounds but also against the irreducible noise from SM
di-Higgs production.

The detector-level distributions shown in
Fig.~\ref{fig:C2HDM_plots2}
display trends similar to those observed for the NMSSM,
although the separation between the signal and the SM
backgrounds is generally more pronounced. 
Among the observables considered here, the
$\tau$-polarisation variable $Z_g$
continues to provide one of the most powerful sources of additional
discrimination, as already shown in the NMSSM analysis.
Finally, 
the invariant mass ratios and angular observables exhibit behaviour
similar to that observed in the NMSSM case, but with a more
pronounced separation between the resonant C2HDM signal and the
non-resonant SM di-Higgs process. 
Together, these observables complement the resonance reconstruction
variables discussed previously and form the basis of the optimised
event selection presented below:

\begin{itemize}

    \item [Pre-Cut:] No. of b-tagged jets, $N_b \geq 2$ and No. of $\tau$-tagged jets, $N_{\tau} \geq 2$.
    \item [Cut A:] $M(bb \tau \tau) \geq 500 ~{\rm GeV}$
    \item [Cut B:] $M(bb)\ /\ M(bb\tau\tau) \leq 0.3,\ M(\tau\tau)\ /\ M(bb\tau\tau) \leq 0.3$
    \item [Cut C:] $p_T(b_1) \geq 90 ~{\rm GeV},\ p_T(\tau_1) \geq 90 ~{\rm GeV}$
    \item [Cut D:] $ 90 ~{\rm GeV} \leq M(bb) \leq 150 ~{\rm GeV}$
    \item [Cut E:] $ 90 ~{\rm GeV} \leq M(\tau\tau) \leq 150 ~{\rm GeV}$
    \item [Cut F:] $\Delta R (bb) \leq 1.5$
    \item [Cut G:] $Z_g \geq 0.27$
\end{itemize}

The detector-level event yields after each successive selection are
summarised in Tab.~\ref{tab:R4_cut_chart2}. The cut-flow clearly
demonstrates the complementarity of the different observables in
isolating the resonant C2HDM signal. The first and most significant
background suppression is achieved through the requirement on the
reconstructed four-body invariant mass,
$M(bb\tau\tau)>500~\mathrm{GeV}$,
which exploits the heavy resonant nature of the signal. This
selection efficiently removes the low-mass $Zjj$ and $t\bar t$
background while retaining a sizeable fraction of both the
SM and C2HDM Higgs-pair events. Within the available
simulated statistics, the QCD multi-jet contribution becomes
negligible after this requirement, confirming that purely
misidentified multi-jet events rarely populate the high-mass signal
region.
The subsequent requirements on the invariant mass ratios and the
transverse momenta of the reconstructed decay products further
enhance the purity of the selected sample. The invariant mass ratio
cuts exploit the strong correlation between the two reconstructed
Higgs candidates and the parent heavy resonance while the
requirements on $p_T(b_1)$ and $p_T(\tau_1)$ take advantage of the
harder kinematic spectra associated with the boosted Higgs bosons in
the C2HDM. The Higgs mass window selections on $M(bb)$ and
$M(\tau\tau)$ then efficiently suppress events that are not
compatible with the decay of two SM Higgs bosons.
The final selections based on the angular separation
$\Delta R(bb)$ and the $\tau$-polarisation observable $Z_g$ provide
additional discrimination against the remaining backgrounds.
Although the $\Delta R$ requirement gives a modest improvement, the
$Z_g$ selection proves particularly powerful. After the complete
selection chain, the reducible SM backgrounds are
essentially eliminated, leaving only the irreducible SM 
di-Higgs production together with the C2HDM signal.

The corresponding statistical significances obtained after each
selection steps are presented in
Tab.~\ref{tab:R4_Significance_table}. Since the purpose of the
analysis is twofold, namely to suppress the dominant SM reducible 
backgrounds while simultaneously identifying the resonant C2HDM
contribution over the irreducible SM di-Higgs background,
we quote the significances for the SM and C2HDM signals
separately, together with the significance associated with the
resonant excess, $S_{\rm signal}$ = $N_{\mathrm{\rm C2HDM}}-N_{\mathrm{\rm SM}}$. Again, a combined ATLAS+CMS predicted significance is also added as $\mathcal{S}^{\rm combined}$.

The statistical significances, computed by using Eq.~(\ref{eq:Significance}), obtained after each stage of the event
selection are summarised in
Tab.~\ref{tab:R4_Significance_table} (up to Cut F). As expected, the significance
improves steadily as the optimised selection progressively suppresses
the dominant SM backgrounds. The largest gain is obtained
through the combined use of the resonance reconstruction variables,
the boosted-event observables and the $\tau$-polarisation-sensitive
quantity $Z_g$, illustrating the complementary nature of these
observables in isolating the resonant C2HDM signal.
An interesting feature of the optimised selection is observed after
the final requirement on the polarisation observable $Z_g$ (Cut G). At this stage, all reducible SM backgrounds considered in the
analysis, namely $t\bar t$, $Zjj$ and QCD multi-jet production,
are removed within the available simulated statistics.
Consequently, the Asimov approximation (~\ref{eq:Significance}),
used in the previous cut stages is no longer applicable,
since the estimated reducible background yield vanishes
within the available Monte Carlo sample.
To account for this, we regularize the null background by adopting a conservative 68.3\% confidence level upper limit of \(B_{\text{eff}} = 1.0\) event. The statistical significance is then evaluated using the exact Poisson probability of observing at least \(N_{\mathrm{o}bs}\) events under this background-only hypothesis,
\begin{equation} 
p_0 = \sum_{n=N_{\rm obs}}^\infty \frac{B_{\text{eff}}^n e^{-B_{\text{eff}}}}{n!}, 
\end{equation}
which is converted into the corresponding one-sided Gaussian significance through \begin{equation}
{\cal S} = \Phi^{-1}(1-p_0), 
\label{eq:Poisson} 
\end{equation} 
where \(\Phi ^{-1}\) denotes the inverse cumulative distribution function of the standard normal distribution. For an expected background of zero (regularized to \(B_{\text{eff}} = 1.0\)), an observation of \(N_{\rm obs} = 22\) events yields a local significance of \(9.7\sigma\), while an observation of \(N_{\rm obs} = 44\) events yields \(15.72\sigma\) (see Tab.~\ref{tab:R4_Significance_table}).

\begin{table}[!t]
    \centering
    \begin{tabular}{c||c|c|c|c|c}
        \toprule
        \multicolumn{6}{c}{
        $\sqrt{s}=14~\TeV,\quad
        \mathcal{L}=4000~\fb^{-1}$} \\
        \toprule
        Dataset
        & SM
        & C2HDM
        & $t\bar t$ lep
        & $t\bar t$ semilep
        & $Zjj$ \\
        \midrule\midrule

        $\sigma~[\fb]$
        & $3.6188$
        & $5.4270$
        & $11493.6$
        & $137921.6$
        & $259895.8$ \\
        \midrule

        Pre-Cut
        & $539$
        & $838$
        & $910630$
        & $2.72\times10^{6}$
        & $737411$ \\
        \hline

        Cut A
        & $97$
        & $196$
        & $86586$
        & $319610$
        & $82474$ \\
        \hline

        Cut B
        & $57$
        & $134$
        & $1622$
        & $11952$
        & $15593$ \\
        \hline

        Cut C
        & $53$
        & $119$
        & $858$
        & $6618$
        & $6236$ \\
        \hline

        Cut D
        & $44$
        & $97$
        & $214$
        & $2390$
        & $1733$ \\
        \hline

        Cut E
        & $29$
        & $57$
        & $106$
        & $0$
        & $347$ \\
        \hline

        Cut F
        & $26$
        & $55$
        & $61$
        & $0$
        & $347$ \\
        \hline

        Cut G
        & $22$
        & $44$
        & $0$
        & $0$
        & $0$ \\

        \bottomrule
    \end{tabular}
    \caption{Cut-flow for the SM and C2HDM di-Higgs signals
    and the dominant backgrounds at the HL-LHC with
    $\sqrt{s}=14~\TeV$ and
    $\mathcal{L}=4000~\fb^{-1}$. Each cut is applied
    sequentially, and the corresponding surviving event yields, $N_L$, are reported. The C2HDM cross section before
    higher-order corrections is taken to be
    $1.5$ times the SM di-Higgs cross section. Higher-order QCD
    corrections are incorporated using
    $K_{hh}=2.2$, $K_{t\bar t}=1.6$, and
    $K_{Zjj}=1.2$. The quoted cross sections and event yields
    include these corrections.}
    \label{tab:R4_cut_chart2}
\end{table}
\begin{table}[!b]
    \centering
    \begin{tabular}{c||c|c|c|c|c|c}
        \toprule
        Cut
        & $\mathcal{S}_{\rm SM}$
        & $\mathcal{S}^{\rm combined}_{\rm SM}$
        & $\mathcal{S}_{\rm C2HDM}$
        & $\mathcal{S}^{\rm combined}_{\rm C2HDM}$
        & $\mathcal{S}_{\rm signal}$
        & $\mathcal{S}^{\rm combined}_{\rm signal}$ \\
        \midrule

        Pre-Cut
        & $0.258$ & $0.365$
        & $0.401$ & $0.567$
        & $0.143$ & $0.202$ \\

        Cut A
        & $0.139$ & $0.196$
        & $0.280$ & $0.396$
        & $0.142$ & $0.200$ \\

        Cut B
        & $0.334$ & $0.472$
        & $0.784$ & $1.109$
        & $0.450$ & $0.637$ \\

        Cut C
        & $0.452$ & $0.640$
        & $1.015$ & $1.435$
        & $0.562$ & $0.795$ \\

        Cut D
        & $0.667$ & $0.943$
        & $1.467$ & $2.075$
        & $0.799$ & $1.130$ \\

        Cut E
        & $1.348$ & $1.907$
        & $2.625$ & $3.712$
        & $1.263$ & $1.787$ \\

        Cut F
        & $1.274$ & $1.802$
        & $2.665$ & $3.769$
        & $1.377$ & $1.947$ \\

        Cut G
        & $9.7$ & $13.71$
        & $15.72$ & $22.23$
        & $4.12$ & $5.83$ \\

        \bottomrule
    \end{tabular}
    \caption{Evolution of the statistical significance ${\cal S}$ from Eq.~(\ref{eq:Significance}) {{(everywhere, except for the first four columns of the last line, where Eq.~(\ref{eq:Poisson}) is used)}}  for the SM and C2HDM processes after each stage of the event selection, for the HL-LHC with $\sqrt{s}=14$ \TeV\ and $\mathcal{L}=4000~{\rm fb}^{-1}$. Here, we treat the SM and C2HDM signals separately before combining them. The quantity $\mathcal{S}_{\rm signal}$ characterises the
    separation between the SM and C2HDM hypotheses, taking
    $S=|N_{\rm C2HDM}-N_{\rm SM}|$ and $B=N_{\rm SM}+N_{\rm bkg}$.We also present the significance  ${\cal S}$ for the ATLAS and CMS datasets combined (in quadrature).}
    \label{tab:R4_Significance_table}
\end{table}

After our selection, we have a non-zero number of events for 
SM di-Higgs production and the resonant C2HDM signal.
Therefore, the experimental challenge is ultimately not background
rejection, rather the discrimination between two different Higgs-pair
production hypotheses. For the BP considered here, the final
event yields are 22 events for the SM and 44 events for the C2HDM (at
$\sqrt{s}=14~\mathrm{TeV}$ with ${\cal L}=$
$4000~\mathrm{fb}^{-1}$). Interpreting the SM prediction
as the reference expectation, the excess associated with the C2HDM
BP corresponds approximately to a (now using the Asimov approximation) significance, 
\begin{equation}
\begin{aligned}
\mathcal{S}_{\rm signal}^{\rm A}
&=
\sqrt{
2\left[
(S+B)\ln\left(1+\frac{S}{B}\right)-S
\right]
}
\simeq 4.12.
\end{aligned}
\end{equation}
i.e., a larger than $4\sigma$ deviation, while the combined significance is more than $6\sigma$. It demonstrates that, once the
reducible backgrounds are eliminated, the remaining sensitivity is
entirely driven by the capability of the detector-level analysis to
distinguish resonant Higgs-pair production in the C2HDM from its irreducible
SM continuum. This highlights the excellent potential of
the $b\bar b\tau^+\tau^-$ channel for probing the C2HDM 
scenario at the HL-LHC.

{Like for the NMSSM case, we conclude this section estimating the reduction factor due to background systematic uncertainties. Using again the significance formula Eq.~\ref{eq:AsimovDB}, the results for the C2HDM BP are shown in Fig.~\ref{fig:C2HDMZDB}. In this case, even considering large systematic uncertainties, the significance after all cuts have been applied stays very large, even in the rather pessimistic assumption that $\Delta B=50\%$.
\begin{figure}[h!]
\centering
\includegraphics[width=0.5\textwidth]{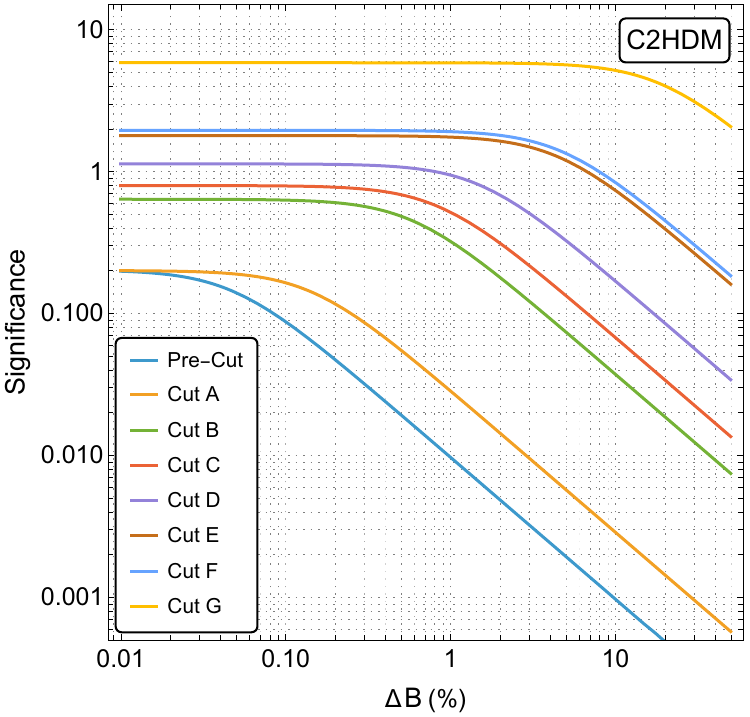}
\caption{\label{fig:C2HDMZDB} Parametric dependence of $\mathcal{S}^{\rm combined}_{\rm signal}$ of the C2HDM scenario on the background systematic uncertainty.}
\end{figure}}


\subsection{NMSSM and C2HDM Comparison}

The detector-level analyses presented above illustrate that, although
both these scenarios predict resonant SM Higgs-pair production and decay into
the $b\bar b\tau^+\tau^-$ final state with the same inclusive rates, their phenomenological
prospects of detection at the HL-LHC are significantly different. In this respect, it is important
to emphasise that the BPs adopted for both BSM scenarios were
not chosen arbitrarily. Rather, they were selected from extensive
scans obtained by imposing all relevant theoretical as well as current
experimental constraints and were extracted from viable regions of parameter space that,
on the one hand, maximise the prospects for observing resonant di-Higgs production in each BSM scenario, and, on the other hand, capture prevalent features of either. 

For the NMSSM BP considered in this work, the effects of the heavy scalar
resonance combined with those of (two) stops in the loop eventually provide only a moderate  difference
with the respect the total SM background (and more so for the reducible than the irreducible noise),
this being due to a large extent to the fact that both the heavy Higgs and stop masses are sub-TeV, thereby trigger a kinematic not dramatically different from the SM one. In fact, 
despite the SM and NMSSM inclusive event rates are notably different and  
the
optimised kinematic selection
efficiently suppresses the $t\bar t$ and $Zjj$ combined noise, the remaining rate for the latter is still large enough to largely obscure the differences between
the SM and NMSSM results. As a consequence, residual  background contributions remain
the limiting factor, leading to an expected statistical significance just 
below the discovery threshold at the HL-LHC.

The situation is markedly different in the C2HDM. Here, the resonant contribution of the heavy scalar occurs for higher masses and so does the non-resonant one due to  the
HTP partners (eight of these) characteristic of CHMs, both having masses beyond the TeV scale, altogether producing then a much stiffer BSM kinematics. Consequently, the C2HDM, even in presence of a phase space suppression due to the large Higgs and HTP masses, offers a larger sensitivity with  respect to the NMSSM case, well into the discover range.
%


\section{Summary and Conclusions}
\label{sec:summa}
In this paper, we have assessed the scope of the HL-LHC in extracting $2b2\tau$  signals of SM di-Higgs production and decay proceeding via a resonant heavy scalar induced by ggF, i.e., $gg\to H\to hh\to b\bar b \tau^+\tau^-$, where $m_H>2m_h$, with the $\tau$'s decaying hadronically. We have done so in the context of two viable theories of the EW scale: Supersymmetry and Compositeness, {for two BPs  yielding}  the same inclusive cross section at the $2b2\tau$ level. The reason for having done so is threefold. Firstly, on the experimental side, SM di-Higgs production (which indeed occurs primarily through ggF) is the highest priority channel to be studied  at the HL-LHC, with a view at not only establishing it as a new Higgs signal but also at accessing the Higgs trilinear self-couplings entering it. Secondly, on the phenomenological side, the choice of the $hh$ decay leading to the $2b2\tau$ final state is dictated by the fact that it is one of the most sensitive probes of BSM interactions being possibly present in 
the SM di-Higgs channel. Thirdly, on the theoretical side, the choice of the two aforementioned BSM frameworks stems from their ability to remedy the hierarchy problem of the SM. In short, the ability to disentangle such a production and decay process from the SM backgrounds, alongside studying its kinematic features, may enable one to shed some light on the actual BSM mechanism responsible for EWSB (which is inserted by hand in the SM).      

The results of our paper, obtained through a realistic MC analysis at detector level (in presence of all dominant backgrounds), are encouraging albeit not definitive. On the one hand, by adopting two representative BPs in the more minimal model realisations of Supersymmetry and Compositeness allowing for the required mass hierarchy (i.e., $m_H>2m_h$), namely, the NMSSM and C2HDM, respectively, we have been able to show some discovery potential of the upgraded CERN machine to the $2b2\tau$ signal in both BSM scenarios,  particularly, when the latest foreseen integrated luminosity of the HL-LHC (i.e., 4000 \fb$^{-1}$) is adopted and ATLAS and CMS datasets are combined: 
this is certainly true for the C2HDM case while the  expected statistical significance for the NMMSM is just below the discovery
threshold. On the other hand, the signals extracted in the NMSSM and C2HDM, when subject to detector effects, do not show the distinctive kinematic features (i.e.,  peaks, thresholds and interferences) 
visible at parton level (described in previous literature of ours, see Refs.~\cite{DeCurtis:2023pus,Moretti:2025dfz}), which would have enabled one to map  these onto the actual underlying theory, whether it be Supersymmetry or Compositeness. 

{The new and dedicated selection strategies used here to extract the NMSSM and C2HDM signals have the following beneficial features: they use the same kinematic variables and the actual cuts are only mildly dependent upon the  resonant mass (i.e., $m_H$) while being robust in extracting the BSM signals with respect to the SM di-Higgs yield.} So, we conclude by advocating their adoption by the multi-purpose experimental collaborations that will be working at the HL-LHC while also trusting their alibity  in further improving these in order to establish enhanced sensitivity to the BSM scenarios considered here, particularly, in controlling background systematics (especially for the NMSSM case). 
{In fact, the two BPs of the NMMSM and  C2HDM put forward here show that  the $2b2\tau$ channel  emerges as a particularly promising probe of such BSM scenarios, in particular, (well) beyond the EW mass scale  where the SM background events can be more effectively cut.}

\acknowledgments
AD is supported by the Carl Trygger Foundation under the project CTS 23:2930.  SM is supported in part through the NExT Institute and STFC Consolidated Grant ST/X000583/1. LP thanks Harri Waltari for useful discussions.
The work of SDC and LDR has been supported by the research grant number 20227S3M3B “Bubble Dynamics in Cosmological Phase Transitions” under the program PRIN 2022 of the Italian Ministero dell’Universit\`{a} e Ricerca (MUR).

\bibliographystyle{JHEP}

\providecommand{\href}[2]{#2}
\end{document}